\documentclass[12pt]{article} 
\pdfoutput=1
\usepackage{graphicx,floatflt,amssymb,rotate}
\usepackage{mathrsfs}
\usepackage{amssymb}
\usepackage{amsfonts}
\usepackage{amsmath}
\usepackage{color}
\usepackage[utf8]{inputenc}
\usepackage{graphicx}
\usepackage{subfig}
\usepackage{multirow}
\usepackage{float}
\usepackage[normalem]{ulem}
\usepackage{pstricks}
\usepackage[toc,page]{appendix}
\usepackage{cite}
\usepackage[colorlinks=true,
linkcolor=red,
urlcolor=blue,
citecolor=blue]{hyperref}
\usepackage{color}
\usepackage[numbers,sort&compress]{natbib}

\def \azeL{{A_0^L}}
\def \azeR{{A_0^R}}
\def \apaL{{A_\parallel^L}}
\def \apaR{{A_\parallel^R}}
\def \apeL{{A_\perp^L}}
\def \apeR{{A_\perp^R}}
\def \re{\text{Re}}

\def \kstar{{K^*}}
\def \eff{{\text{eff}}}

\def\R1{\varepsilon_1}
\def\E8{\varepsilon_8}

\newcommand{\bea}{\begin{eqnarray}}
\newcommand{\eea}{\end{eqnarray}}
\newcommand{\bd}{\begin{displaymath}}
\newcommand{\ed}{\end{displaymath}}

\newcommand{\be}{\begin{equation}}
\newcommand{\ee}{\end{equation}}
\newcommand{\bi}{\begin{itemize}}
\newcommand{\ei}{\end{itemize}}
\newcommand{\ord}{{\cal O}}

\DeclareUnicodeCharacter{2212}{-}
\begin{document}
\vskip 30pt  
 
\begin{center}  
{\Large \bf Reappraisal of the minimal flavoured $Z^{\prime}$ scenario} \\
\vspace*{1cm}  
\renewcommand{\thefootnote}{\fnsymbol{footnote}}  
{{\sf ~Tirtha Sankar Ray\footnote{email: tirthasankar.ray@gmail.com}}, 
{\sf ~Avirup Shaw\footnote{email: avirup.cu@gmail.com}}
}\\
\vspace{10pt}  
{{\em Department of Physics, Indian Institute of Technology Kharagpur, Kharagpur 721302, India}}
\normalsize  
\end{center} 

\begin{abstract}
\noindent
Recent results from the intensity frontier indicate the tantalizing possibility of violation in lepton flavour universality. In light of this we revisit the minimal phenomenological $Z'$ model taking in account both vectorial and axial-vectorial flavour violating couplings to the charged leptons. We make a systematic study to identify the minimal framework that can simultaneously explain the recent results on anomalous magnetic moment of muon and electron while remaining in consonance with $R_{K^{(*)}}$, $B^0_s-\bar{B^0_s}$ mixing and angular observables in the $B^+\to K^{+*} \mu^+\mu^-$ channel reported by the LHCb collaboration. We demonstrate that the neutrino trident data imply a further ${\rm SU(2)}_L$ violation in the leptonic couplings of the exotic $Z'$.

\vskip 5pt \noindent 
\end{abstract}

\setcounter{footnote}{0}  
\renewcommand{\thefootnote}{\arabic{footnote}}

\section{Introduction}\label{intro}
With continued improvement in resolution, the consistency of anomalous results from the intensity frontier may be the most significant indication of Beyond Standard Model (BSM) physics that we have today in hard experimental data.

Recently the measurement of the $(g-2)_\mu$ at the Fermi National Accelerator Laboratory (FNAL) \cite{Muong-2:2021ojo} has added to this intrigue. The dramatic improvement in the resolution in the recent results has pushed the deviation from Standard Model (SM) prediction at $4.2~\sigma$ with $\Delta a_{\mu} = a_\mu^{\rm exp} - a_\mu^{\rm SM} \sim (251 \pm 59)\times 10^{-11}.$ Notwithstanding the recent lattice results \cite{Borsanyi:2020mff} this deviation has withstood scrutiny for some time now. Interestingly if one compares this with the status of $(g-2)_e$ measurement based on the Lawrence Berkeley National Laboratory (LBNL) determination of the structure constant based on cesium \cite{Parker:2018vye} one obtains a moderate deviation from SM at $2.4~\sigma,$ with the opposite pull at $\Delta a_e  \sim (-8.8 \pm 3.6)\times 10^{-13}$ \cite{Hanneke:2008tm}. This sign flip if true cannot be explained by simple mass scaling $\Delta a_e /\Delta a_{\mu} \sim m_e^2/m_\mu^2$ and should be construed as an indication of Lepton Flavour Universality Violation (LFUV) of any underlying New Physics (NP).

One can trace the imprint of this violation of lepton flavour universality in the rare decays of the $B$-mesons giving a more compelling experimental basis for the LFUV hypothesis. Indications of such LFUV can be seen in the flavour changing neutral current (FCNC) induced decays involving the $b\to sl^+l^-$ transitions. These are indeed easy picking ground in flavour physics searches for NP owing to the Glashow Iliopoulos Maiani (GIM) suppression of tree-level contribution to them within SM. In this context the observable $R_{K^{(*)}}$ is of significance as they are relatively independent of the form factor uncertainties \cite{SHIFMAN1979385, Colangelo:2000dp}. SM predictions \cite{Descotes-Genon:2015uva,Bordone:2016gaq,Capdevila:2017bsm} of these parameters exhibit $2-3~\sigma $ deviations from the corresponding experimental results reported by LHCb \cite{LHCb:2021trn, Aaij:2017vbb}. Tying nicely with the paradigm of LFUV in underlying physics. Thus it is not surprising that there exists extensive literature that study various motivated BSM framework that tries to explain these anomalous results individually or in combinations \cite{Gauld:2013qba,Glashow:2014iga,Bhattacharya:2014wla, Crivellin:2015mga, Crivellin:2015era,Celis:2015ara,Sierra:2015fma,Belanger:2015nma,Gripaios:2015gra, Allanach:2015gkd,Fuyuto:2015gmk,Chiang:2016qov,Boucenna:2016wpr,Boucenna:2016qad,Celis:2016ayl, Altmannshofer:2016jzy,Bhattacharya:2016mcc,Crivellin:2016ejn,Becirevic:2016zri,GarciaGarcia:2016nvr,Bhatia:2017tgo,Ko:2017yrd,Chen:2017usq,Baek:2017sew, King:2017anf, King:2018fcg, Dasgupta:2018nzt, Biswas:2019twf, Dwivedi:2019uqd, CarcamoHernandez:2019ydc, Bodas:2021fsy, Biswas:2021dan, Hiller:2014yaa,Biswas:2014gga,Gripaios:2014tna,Sahoo:2015wya,Becirevic:2015asa, Alonso:2015sja,Calibbi:2015kma, Huang:2015vpt,Pas:2015hca,Bauer:2015knc,Fajfer:2015ycq,Barbieri:2015yvd, Sahoo:2015pzk, Dorsner:2016wpm,Sahoo:2016nvx,Das:2016vkr,Chen:2016dip,Becirevic:2016oho,Becirevic:2016yqi,Sahoo:2016pet,Barbieri:2016las,Cox:2016epl, Ma:2001md, Baek:2001kca, Heeck:2011wj, Harigaya:2013twa, Altmannshofer:2016brv, Biswas:2016yan, Biswas:2016yjr, Banerjee:2018eaf, Huang:2020ris, Dinh:2020inx, Chakraborti:2021kkr, Chakraborti:2021dli}. Moreover, after the declaration of the updated results \cite{LHCb:2021trn, Muong-2:2021ojo}, several authors and collaboration published various articles in different existing and/or new BSM scenarios. Among the different categories of BSM frameworks, a kind of scenario with a non-standard neutral massive boson ($Z'$) is very popular and effective, for the combined explanation of $R_K$ (including other $b\to s ll$ anomalies) and $(g-2)_\mu$ \cite{Arcadi:2021cwg, Alvarado:2021nxy, Davighi:2021oel, Darme:2021qzw, Lee:2021ndf, Greljo:2021npi,Wang:2021uqz, Navarro:2021sfb, Bause:2021prv, Ko:2021lpx}. 

In the present paper we revisit one of the simplest of these framework that extends the SM to include an additional massive abelian gauge boson $Z'$ with flavour violating couplings. With Occum's razor in hand we build bottom up model of the $Z'$ guided solely by the requirement of the experimental data remaining agnostic to any UV completion. Our approach is to pare down to the minimal version of the model that would be simultaneously in consonance with the various experimental results at low energy intensity frontiers with focus on observables that indicate LFUV. Our notion of simplicity will be guided by economy of new parameters rather than any symmetry or embedding considerations. Starting with the simplest two parameter model of $Z'$ mass and universal couplings, we systematically build this  phenomenological model of $Z'$ from bottom up by adding new parameters as the experimental data demand the same. In this context, we demonstrate that the simultaneous inclusion of flavour violating vectorial and axial-vectorial couplings to the leptons is a prudent choice in constructing such minimal models. 

Once we establish a \textit{data driven}  minimal phenomenological $Z'$ model for LFUV we explore the region of parameter space of such a setup that would be in agreement with relevant experimental results  including $B^0_s - \bar{B}^0_s$ mixing \cite{Zyla:2020zbs}, recent LHCb results of angular observables in the $B^+ \to K^{+*}\mu^+\mu^-$ channel \cite{LHCb:2020gog} and some other constraints that are related to $b\to s$ transitions \cite {LHCb:2020pcv, LHCb:2013pra, LHCb:2014vgu, Belle:2017oht}. We have also taken into account the constraints from collider and fixed target experiments \cite{Zyla:2020zbs, Mishra:1991bv}. 

The article is organised as follows. In the Sec.~\ref{model} we explore different effective $Z'$ scenarios in the context of recent data of $(g-2)_\mu$ and $(g-2)_e$. Then in Sec.~\ref{btosll_ano} we resolve different anomalies related to $b\to s ll$ transitions in a particular economic effective $Z'$ scenario. After that in Sec.~\ref{angular} we discuss the new results of LHCb related to angular observables of the decay $B^+\to K^{+*}\mu^+\mu^-$. We discuss our numerical results in Sec.~\ref{neu_res}. Then in  Sec.~\ref{excons} we comment on some relevant constraints. In Sec.~\ref{plus_g2e} we re-examine the scenarios that we consider in this article in the context of latest experimental value of $(g-2)_e$ for which the $\Delta a_e$ is positive. Finally, in Sec.~\ref{concl} we will conclude our findings.

\section{Resolution of anomalous magnetic moment of charged lepton in minimal $Z'$ scenario(s)}\label{model}
 The magnetic moment $\vec{\mathbb{M}}$ of a charged lepton ($l$) can easily be defined in terms of its spin $\vec{\mathbb{S}}$ and gyromagnetic ratio ($g_{l}$) using the Dirac equation as follows
\begin{eqnarray}
\vec{\mathbb{M}}= g_{l} \dfrac{e}{2\,m_l} \vec{\mathbb{S}}\,.
\label{mug2}
\end{eqnarray}
Ideally the value of $g_{l}$ is equal to 2. Quantum correction provides marginal shift within the SM. In order to estimate this deviation of $g_{l}$ from its tree-level value one can usually define the parameter,
\begin{eqnarray}
a_{l} = \dfrac{g_{l}-2}{2}\,.
\end{eqnarray}

\begin{figure}[H]
\begin{center}
\includegraphics[height=4.5cm,width=5cm,angle=0]{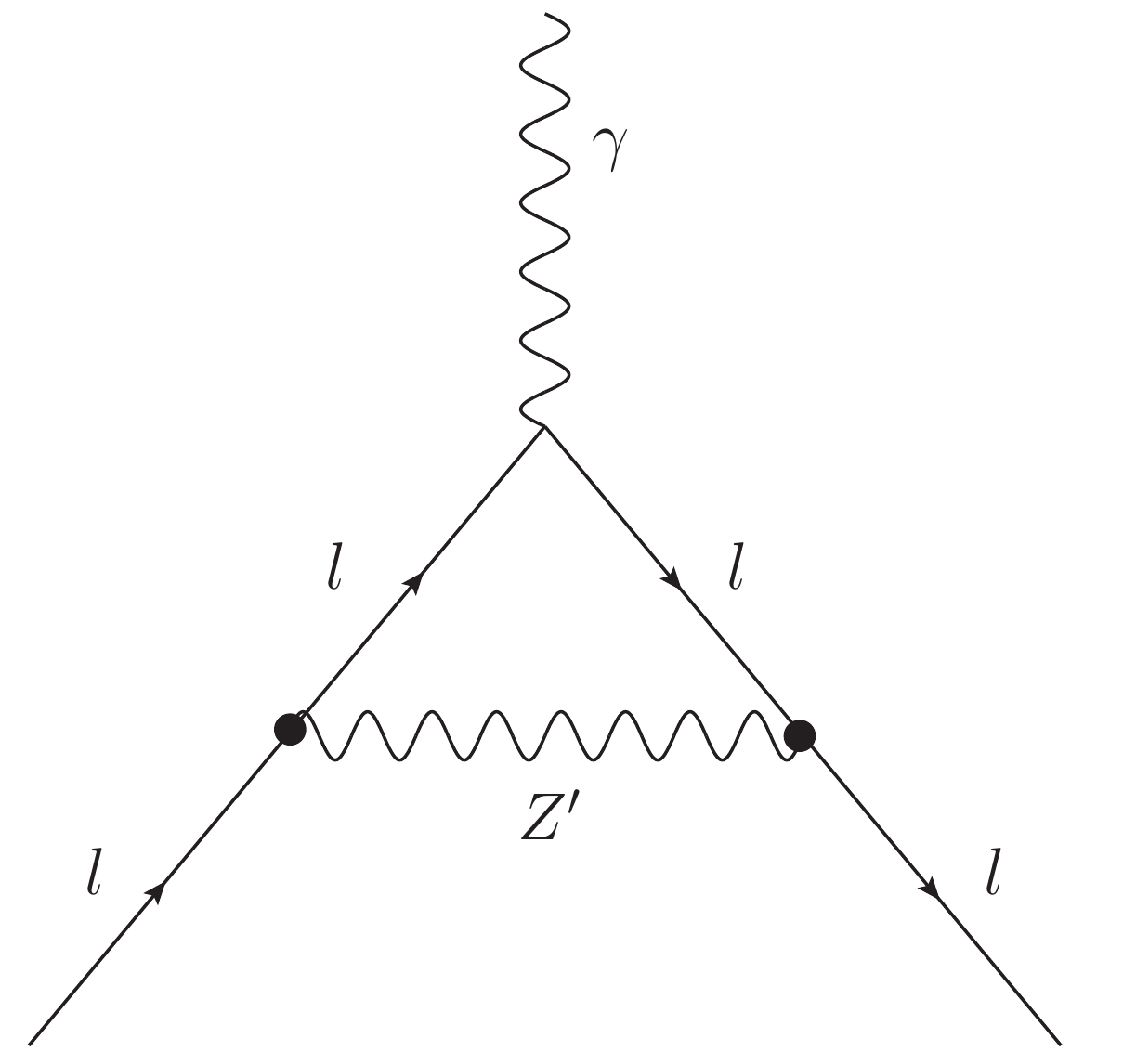}
\caption{Relevant Feynman diagram that is contributed to the $(g-2)_l$ in addition to the SM.}
\label{muong2}
\end{center}
\end{figure}
We consider the extension of the SM incorporating a massive $Z'$ as mass $M_{Z'}$ having flavoured vectorial and axial-vectorial couplings as given by a generic effective Lagrangian,
\be\label{aZp_bZp_same}
 \mathcal{L}\in \bar{l}\gamma^\alpha(a^l_{Z^{\prime}}+\gamma^5 b^l_{Z^{\prime}})l~Z'_{\alpha}\,,
\ee
where, $a^l_{Z^{\prime}}$ and $b^l_{Z^{\prime}}$ (where $l\in e, \mu$) are the vectorial and axial-vectorial couplings of light charged leptons with $Z'$ boson. This leads to an additional contribution to the anomalous magnetic moment of charged lepton depicted in Fig.~\ref{muong2} and the corresponding contribution is given by \cite{Gninenko:2001hx, Baek:2001kca, Biswas:2019twf, Biswas:2021dan},

\begin{equation}\label{g2_Z}
\Delta a_l^{Z'} = \frac{1}{8\pi^2}~
\left((a^l_{Z'})^2F_{a^l_{Z^{\prime}}}(R_{Z'})-(b^l_{Z'})^2F_{b^l_{Z^{\prime}}}(R_{Z'}))\right)\,,\\
\end{equation}
with $R_{Z'}\equiv M^2_{Z'}/m^2_{l}$. Here, $M_{Z'}$ is the mass of $Z'$ boson and $m_l$ is the mass of charged lepton. The loop functions corresponding to vectorial and axial-vectorial interactions are given as follows, 
\begin{eqnarray}
F_{a^l_{Z^{\prime}}}(R_{Z'}) &=& \int_0^1 dx\, \frac{2x(1-x)^2}{(1
 -x)^2+R_{Z'} x}
 \;, \\
F_{b^l_{Z^{\prime}}}(R_{Z'}) &=& \int_0^1 dx\,\frac{2x(1-x)(3+x)}{(1
 -x)^2+R_{Z'} x}\;.
\end{eqnarray}

\subsection{The Minimal Model}
We now embark on identifying the minimal flavoured model of $Z'$ that can explain the anomalous magnetic moment of the charged leptons. Our hunt will be guided by the recent experimental data on the anomalous magnetic moment of the charged leptons that we now briefly summarise:
\begin{itemize}
\item In the case of muon ($\mu$), over the last two decades there is an enduring deviation ($\gtrapprox 3.5 \sigma$) between theoretical prediction and the corresponding experimental data. Recently, a measurement of the $(g-2)_\mu$ has been reported by FNAL \cite{Muong-2:2021ojo},
\begin{eqnarray}
a_{\mu}^{\rm exp} = 116592061\pm 41 \times 10^{-11}\,.
\label{mug2exp}
\end{eqnarray}
The state of the art SM theoretical prediction is \cite{Aoyama:2020ynm}
\begin{eqnarray}
a_{\mu}^{\rm SM} = 116591810\pm 43 \times 10^{-11}\,.
\label{mug2th}
\end{eqnarray}
This amounts to a disagreement between SM prediction and experiment at $4.2 \sigma$ parametrised by,
\begin{eqnarray}
\Delta a_{\mu} = a_{\mu}^{\rm exp} - a_{\mu}^{\rm SM}
=(251\pm 59) \times 10^{-11}\,.
\label{mug2delta}
\end{eqnarray}
\item Similarly there exists a milder disagreement if we compare theoretical and experimental values of $(g-2)$ in the electronic sector. For example, SM prediction of $(g-2)_e$ \cite{Aoyama:2017uqe} was determined with the value of fine structure constant, evaluated at the Berkeley laboratory by performing high precision measurement using cesium atoms~\cite{Parker:2018vye}. This value moderately deviates ($\simeq 2.4\sigma$) from the corresponding experimental result \cite{Hanneke:2008tm} and is parametrised as,
\begin{equation}
\Delta a_e = a_e^{\rm exp} - a_e^{\rm SM} =(-8.8\pm 3.6)\times 10^{-13}\,.
\label{eg2negative}
\end{equation}
Interestingly the deviation in the anomalous magnetic moment for the electronic and muonic sector have opposite sign. It is difficult to account for this from a simple mass scaling ($\Delta a_e / \Delta a_\mu \sim m^2_e/m^2_\mu\sim 10^{-5}$). This can be construed as an  evidence of lepton flavour non-universality of any underlying NP\footnote{For an explanation of the anomalous magnetic moment of leptons within lepton flavour universality see \cite{Hiller:2019mou}.}. In this paper we systematically study the minimal flavoured $Z'$ model that can simultaneously explain the deviation in the anomalous magnetic moment in the electronic and muonic sector.
\end{itemize}
\begin{enumerate}
\item First we consider that the $Z'$ has universal vectorial interaction with charged lepton with the effective Lagrangian of the form,
\be\label{aZp_mu1}
 \mathcal{L}\in \bar{l}\gamma^\alpha(a_{Z^{\prime}})l~Z'_{\alpha}\,,
\ee
where, $l\in (e,\mu)$. This scenario has two free parameters $M_{Z^{\prime}}$ and $a_{Z^{\prime}}$. With this setup, while one can tune $a_{Z^{\prime}}$ to explain the result of anomalous magnetic moment of muon. However, there is no possibility to explain the relative sign difference between the two leptonic generation.


\item We now consider non-zero vectorial ($a_{Z^{\prime}}$) and axial-vectorial ($b_{Z^{\prime}}$) couplings with both muon and electron. The effective interaction is given by,
\be\label{aZp_bZp_same1}
 \mathcal{L}\in \bar{l}\gamma^\alpha(a_{Z^{\prime}}+\gamma^5 b_{Z^{\prime}})l~Z'_{\alpha}.
\ee
This extends the number of free parameters to three viz $a_{Z^{\prime}}$, $b_{Z^{\prime}}$ and $M_{Z^{\prime}}$. This is still unable to simultaneously explain the measured value of $\Delta a_\mu$ and $\Delta a_e$. The region of the parameter space allowed by $\Delta a_\mu$ and $\Delta a_e$ individually are shown in Fig.~\ref{fig:aZbZ} in $a_{Z^{\prime}}$ vs $b_{Z^{\prime}}$ plane for different values of $M_{Z'}$ in the range [$10^{-3}$ to $10^3$] GeV. We obtain no overlap at al. 


\begin{figure}[t!]
\begin{center}
\subfloat[]{\label{fig:aZbZ}\includegraphics[height=7.5cm,width=8.5cm,angle=0]{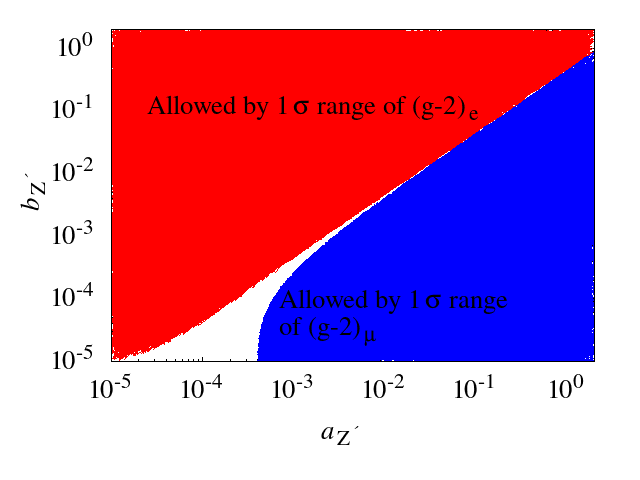}}
\subfloat[]{\label{fig:e_mu_a0e_b0mu}\includegraphics[height=7.5cm,width=9.5cm,angle=0]{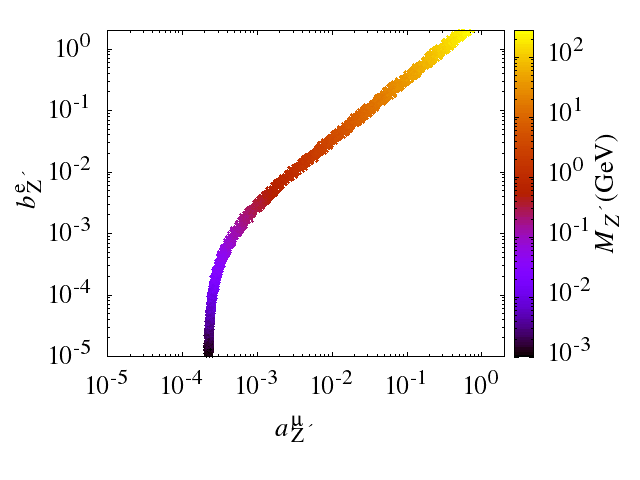}}
\caption{Left panel: 1$\sigma$ allowed parameter space by $(g-2)_\mu$ (blue) and $(g-2)_e$ (red) in $a_{Z^{\prime}}$ and $b_{Z^{\prime}}$ plane. Right panel: 1$\sigma$ allowed parameter space in $a^\mu_{Z^{\prime}}$ vs $b^e_{Z^{\prime}}$ plane with the allowed values of $M_{Z^{\prime}}$ (in GeV).} 
\end{center}
\end{figure} 

\item Keeping with three parameter models we now explore the possibility of addressing the anomalous magnetic moments using an interaction of the following form,
\be\label{aZp_mu}
 \mathcal{L}\in \bar{\mu}\gamma^\alpha(a^\mu_{Z^{\prime}})\mu~Z'_{\alpha}+\bar{e}\gamma^\alpha(\gamma^5 b^e_{Z^{\prime}})e~Z'_{\alpha}\,.
\ee
With these combination one can simultaneously explain the $(g-2)_{\mu}$ and $(g-2)_{e}$ data at al. We have shown the allowed parameter space in $a^\mu_{Z^{\prime}}$ vs $b^e_{Z^{\prime}}$ plane while the values of $M_{Z^{\prime}}$ are shown by colour codes. Expectedly with increasing values of $a^\mu_{Z^{\prime}}$ and $b^e_{Z^{\prime}}$ a large $M_{Z^{\prime}}$ is required to reduce the loop effects by suppressing the propagator as can be read off from Fig.~\ref{fig:e_mu_a0e_b0mu}. 
\end{enumerate}
 
\section{LFUV From \boldmath$B$-sector}\label{btosll_ano}
A synergy of experimental results in the decay of $B$-meson provide further credence to the emergent paradigm of LFUV. As for example $R_{K^{(*)}}$ define as \cite{Hiller:2003js},
\begin{align}\label{Rth}
R_{K^{(*)}} \equiv \frac{\int^{q^2_{\rm max}}_{q^2_{\rm min}} \frac{d\Gamma\left({B} \rightarrow K^{(*)}
 \mu^+ \mu^-\right)}{d q^2} d q^2}
{\int^{q^2_{\rm max}}_{q^2_{\rm min}} 
 \frac{d\Gamma\left({B} \rightarrow K^{(*)} e^+ e^-\right)}{d q^2} d q^2}\;,
\end{align}
where $q^2$ represents the dilepton mass squared with the limits $q^2_{\rm max}= (m_B-m_{K^{(*)}})^2$, $q^2_{\rm min}= 4m^2_l$ and $m_B$ represents the mass of $B$-meson. When QED corrections are included in the SM, these ratios are close to modulo one \cite{Isidori:2020acz}. We summarise some of the relevant FCNC related experimental parameters in Table \ref{rkrkstdata}.  The imprint of NP in this data is becoming increasingly apparent.
\begin{table}[!h]
\centering
\begin{tabular}{|c|cr|cr|cr|} 
\hline 
Observable & SM prediction &  & Measurement  &  & Deviations &\\
\hline 
$R_K : q^2 = [1.1,6] \, \text{GeV}^2$  & $1.00 \pm 0.01 $&  \cite{Descotes-Genon:2015uva,Bordone:2016gaq} &  $0.846^{+0.042+0.013}_{-0.039-0.012}$  & \cite{LHCb:2021trn} &  3.1$\sigma$ &\\
\hline
$R_{K^*} ^{\rm low}: q^2 = [0.045,1.1] \, \text{GeV}^2$  & $0.92 \pm 0.02$  &  \cite{Capdevila:2017bsm} &  $0.660^{+0.110}_{-0.070} \pm 0.024$  & 
\cite{Aaij:2017vbb} & $2.1\sigma-2.3\sigma$ &\\
\hline
$R_{K^*}^{\rm central} : q^2 = [1.1,6] \, \text{GeV}^2$  & $1.00 \pm 0.01 $&  \cite{Descotes-Genon:2015uva,Bordone:2016gaq} &  $0.685^{+0.113}_{-0.069} \pm 0.047$  &  \cite{Aaij:2017vbb} & $2.4\sigma-2.5\sigma$ &\\
\hline
\end{tabular}
\caption{The experimental values of $R_K$ and $R_{K^*}$ along with their SM predictions for different ranges of $q^2$.}
\label{rkrkstdata}
\end{table} 
Considering this, we now proceed to check the compatibility of the minimal $Z'$ model discussed in the previous section with the experimental data related to $b\to sll$ transitions. With this non-universal nature of the coupling of $Z'$ to $e$ and $\mu$ it is expected that the decay widths for $b\to s\mu^+\mu^-$ and $b\to s e^+e^-$ will be different. The possibility to exploit this property to resolve the above mentioned LFUV anomalies is optimistic. In order to couple to the quark sector we introduce a single new flavour off diagonal interaction with coupling $\textsl{g}_{bs}$ 
\be\label{btos_int}
 \Delta\mathcal{L}\in \textsl{g}_{bs}(\bar b \gamma^\alpha P_L s) Z'_{\alpha}\,,\ee
in addition to the Lagrangian given in Eq.~\ref{aZp_mu}. In the rest of this paper, we refer to this minimally flavoured scenario as MFS. The effective interaction between $bsZ'$ imply new contribution to   the ${B^0_s}-\bar{B^0_s}$ oscillation at tree-level. This can provide stringent constraint in the parameter space and thus we incorporate the constraint of ${B^0_s}-\bar{B^0_s}$ oscillation in our analysis.

The analysis that follows our approach is to reconstruct the flavour observables using their proper definition within MFS and implement them in our numerical analysis. This may be {\it contrasted} with the approach where fit values of Wilson Coefficients (WCs) are utlised.

\subsection{The \boldmath$B\to K^{(*)}l^+l^-$ transition} 
\begin{figure}[t!]
\begin{center}
\includegraphics[height=3.5cm,width=7cm,angle=0]{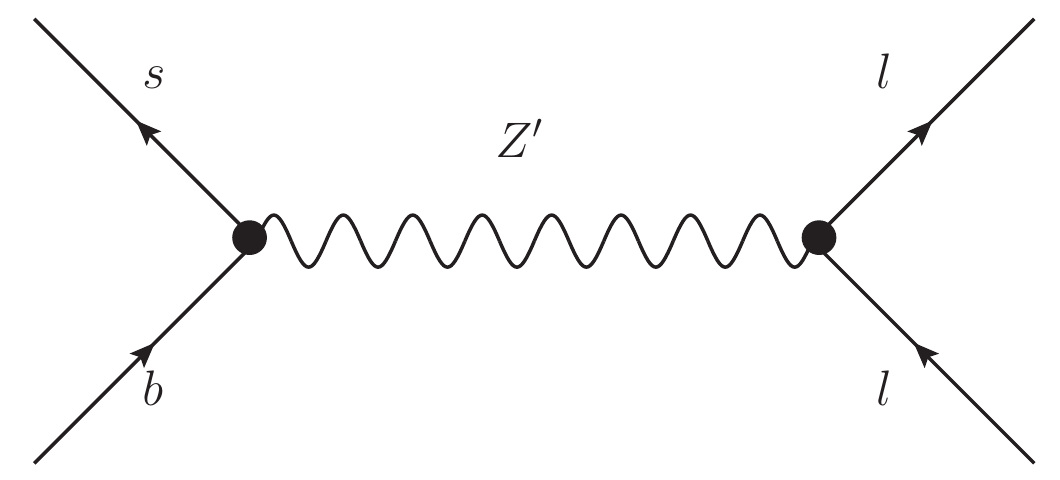}
\caption{Tree-level Feynman diagram that contributes to $b\to sl^+l^-$ transition mediated by $Z'$ boson.} 
\label{Zptree}
\end{center}
\end{figure}

The hadronic decay $B\to K^{(*)}l^+l^-$ is driven by the underlying transition $b\to sl^+\l^-$ at the quark level. The effective Hamiltonian at hadronic scale $Q\sim m_b$ is given by \cite{Buchalla:1995vs}
\begin{equation} 
\centering
{\cal H}_{\rm eff}(b\to s l^+l^-) =
{\cal H}_{\rm eff}(b\to s\gamma)  - \frac{G_{\rm F}}{\sqrt{2}}\frac{\alpha_{\rm em}}{\pi} V_{ts}^* V_{tb}
\left[ C_{9}(Q) \mathcal{O}_{9}+
C_{10}(Q) \mathcal{O}_{10}    \right]\,,
\label{Heff2_at_mu}
\end{equation}

where,
\begin{equation}\label{Q9V}
\mathcal{O}_{9}    = (\bar{s}\gamma^\alpha P_L b)  (\bar{l}\gamma_\alpha l)\,,         
\qquad
\mathcal{O}_{10}  =  (\bar{s}\gamma^\alpha P_L b) (\bar{l}\gamma_\alpha\gamma_5 l)\,,
\end{equation}
are the two crucial operators for this transition. In the first part of the Hamiltonian (see Eq.~\ref{Heff2_at_mu}) there is no NP contribution, while in the remaining part, along with the SM contribution there is NP contribution from tree-level $Z'$ exchange (see Fig.~\ref{Zptree}). Since $Z'$ is assumed to couple only to left handed SM fermion, therefore the  corresponding chirality flipped operators are not generated. The relevant WCs corresponding to the operators $\mathcal{O}_{9}$ and $\mathcal{O}_{10}$ 
have contained the tree-level NP contributions from a phenomenological $Z'$ define in Eqs.~\ref{aZp_mu} and \ref{btos_int} are given by \cite{Buras:2012dp},
\begin{eqnarray} 
C^{\rm NP}_{9} &=& -\frac{\sqrt{2}\pi}{G_F \alpha_\text{em} V_{ts}^* V_{tb}} \frac{\textsl{g}_{bs} a^\mu_{Z^{\prime}}}{M^2_{Z^{\prime}}}\,,\label{C9NP} \\
C^{\rm NP}_{10} &=& -\frac{\sqrt{2}\pi}{G_F \alpha_\text{em} V_{ts}^* V_{tb}}\frac{\textsl{g}_{bs} b^e_{Z^{\prime}}}{M^2_{Z^{\prime}}}\,,
\label{C10NP}
\end{eqnarray}
where $G_F$ is the Fermi constant, $\alpha_\text{em}$ represents the fine structure constant and $V_{ij}$ stands for the Cabibbo Kobayashi Maskawa (CKM) matrix elements. 
\subsubsection{\boldmath$R_K$}\label{ddwk}
In light of the recent result on the differential distribution of the $B^+\to K^+l^+l^-$ transitions from LHCb \cite{LHCb:2021trn} in this section we briefly discuss the decay distribution of this transition. The differential branching fraction for $B^+\to K^+l^+l^-$ is written as \cite{Altmannshofer:2014rta}
\begin{align}\label{btokdw}
\frac{d\Gamma(B^+\to K^+l^+l^-)}{dq^2} = \frac{G_F^2\alpha^2_\text{em}|V_{tb}V_{ts}^*|^2}{2^{10} \pi^5 m_B^3} \lambda^{3/2}(m_B^2,m_{K}^2,q^2)
\left( |F_V|^2+|F_A|^2 \right),
\end{align}
where,
\begin{align}
\lambda(a,b,c)&= a^2  +b^2 + c^2 - 2 (ab+ bc + ac)
\,,\\
F_V(q^2) &= C_{9}^\text{eff}(q^2) f_+(q^2)
+ \frac{2m_b}{m_B+m_K}C_{7}^\text{eff} f_T(q^2)
\,,
\\
F_A(q^2) &= C_{10}f_+(q^2)
\,.
\end{align}
For completeness we have given the relevant expressions for the WCs in Appendix \ref{NDR}. $f_+$ and $f_T$ are the relevant form factors. Corresponding details are given in the Appendix~\ref{sec:ftfprel}. Using the Eqs.~\ref{btokdw} and \ref{Rth} we can evaluate the $R_K$.

\subsubsection{\boldmath$R_{K^{*}}$}\label{ddwkst}
Similar to the earlier section, in order to determine the $R_{K^{*}}$ here we would like to calculate the differential decay rate of $B\to K^{0*}l^+l^-$ with respect to $q^2$ and which is given as follows \cite{Bobeth:2008ij,Matias:2012xw}
\begin{eqnarray}
	\label{ddwrkst}
	\frac{d\Gamma(B^0\to K^{0*}l^+l^-)}{dq^2}&=&\frac{1}{4}(3I_1^c+6I_1^s-I_2^c-2I_2^s)\;.
\end{eqnarray}
Using the above expression and Eq.~\ref{Rth} we can determine the observable $R_{K^{*}}$. The angular coefficients $I^{c,s}_{1,2}$ that are involved in the above Eq.~\ref{ddwrkst} can be defined as follows \cite{Altmannshofer:2008dz}
\begin{eqnarray}
  I_1^s  &=& \frac{(2+\beta_l^2)}{4} \left[|\apeL|^2 + |\apaL|^2 + (L\to R) \right] 
            + \frac{4 m_l^2}{q^2} \re\left(\apeL^{}\apeR^* + \apaL^{}\apaR^*\right), \label{I1s} \\
  I_1^c  &=&  |\azeL|^2 +|\azeR|^2  + \frac{4m_l^2}{q^2} 
               \left[|A_t|^2 + 2\re(\azeL^{}\azeR^*) \right],\label{I1c}\\ 
  I_2^s  &=& \frac{ \beta_l^2}{4}\left[ |\apeL|^2+ |\apaL|^2 + (L\to R)\right], \label{I2s}\\
  I_2^c  &=& - \beta_l^2\left[|\azeL|^2 + (L\to R)\right],
\label{I2c}
\end{eqnarray}
with $\beta_l=\sqrt{1-4m_l^2/q^2}$. The functions $A_{L/R,i}$ are called the transversity amplitudes that can be decomposed in terms of appropriate WCs and form factors. Details of the relevant form factors are provided in Appendix \ref{BSZ:ff}. The WCs ($C_9$ and $C_{10}$) are containing the NP contributions from $Z'$ exchange (see Eqs.~\ref{C9NP} and \ref{C10NP}). The complete expressions for the transversity amplitudes have been given in Appendix \ref{trvnampli}.

\subsection{Constraint from ${B^0_s}-\bar{B^0_s}$ mixing due to tree-level contributions of $Z^\prime$}\label{bbar_con}
In this section we discuss the constraint from the mass difference ($\Delta M_s$) between the ${B^0_s}$ meson mass eigenstates arising from the ${B^0_s}-\bar{B^0_s}$ mixing phenomena. The SM contribution to this $\Delta B=2$ transition process through the top quark mediated box-diagram \cite{Buras:1990fn, Urban:1997gw} and is numerically given as \cite{Lenz:2019lvd} $(\Delta M_s)_{\rm SM}=(18.77\pm 0.86){\rm ps}^{-1}$ that is in good agreement with experimental value \cite{Zyla:2020zbs} $(\Delta M_s)_{\rm exp}=(17.749\pm 0.019\pm 0.007){\rm ps}^{-1}$.
\begin{figure}[htbp!]
\begin{center}
\includegraphics[height=3.5cm,width=7cm,angle=0]{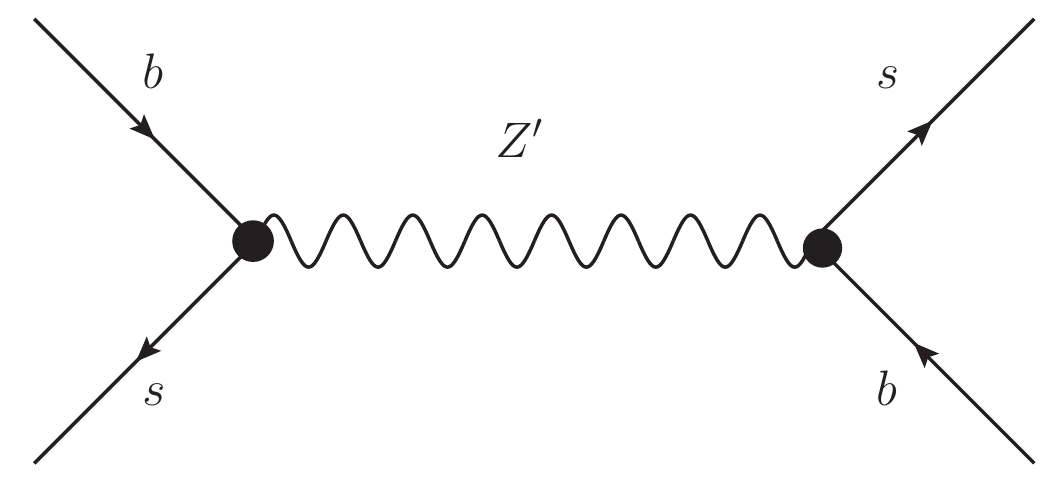}
\caption{Tree-level Feynman diagram that contributes to ${B^0_s}-\bar{B^0_s}$ transition mediated by $Z'$ boson.} 
\label{bs_bs}
\end{center}
\end{figure}
In the minimal flavoured $Z'$ model there exists a tree-level $Z'$ contribution depicted in Fig.~\ref{bs_bs} and can be defined as \cite{Buras:2012dp}
\be\label{Zprime1}
\Delta M_s(Z^\prime)=
\left[\frac{g_{bs}}{V^*_{ts}V_{tb}}\right]^2
\frac{4\tilde r \pi \sin^2\theta_W}{\sqrt{2}G_F\alpha_\text{em} M^2_{Z^\prime}}, 
\ee
with
\be
\tilde r=\frac{C_1^{\rm VLL}(M_{Z^\prime})}{0.985}
             \eta_6^{6/21}\left[1+1.371\frac{\alpha_s^{(6)}(m_t)}{4\pi}(1-\eta_6)\right],
\label{rtilde}
\ee
where,
\be\label{equ:WilsonZ}
C_1^\text{VLL}(Q)=
1+\frac{\alpha_s}{4\pi}\left(-2\log\frac{M_{Z'}^2}{Q^2}+\frac{11}{3}\right)\;.
\ee
The above quantity depicts $\ord(\alpha_s)$ QCD corrections to $Z'$ tree-level exchange
\cite{Buras:2012fs}
and
the two factors containing
\be
\eta_6=\frac{\alpha_s^{(6)}(M_{Z^\prime})}{\alpha_s^{(6)}(m_t)}\;,
\ee
designate together NLO QCD renormalisation group evolution from top quark mass ($m_t$) to
$M_{Z^\prime}$ as given in \cite{Buras:2001ra}. In our scan we restrict the value of ${(\Delta M_s)_{\rm exp}}/{(\Delta M_s)_{\rm SM}}$ within the 2 $\sigma$ allowed range $(0.946\pm 0.086)$.

\section{Angular observables for \boldmath$B^+\to K^{+*}\mu^+\mu^-$ transition}\label{angular}
The LHCb collaboration recently reported results for angular observables of the $B^0\to K^{0*}\mu^+\mu^-$ channel \cite{LHCb:2020lmf}. Interestingly, the observation by LHCb collaboration~\cite{LHCb:2020lmf} on the $B^0\to K^{0*}\mu^+\mu^-$ decay channel is in tension with respect to the SM prediction of observable $P'_5$. Similarly, a latest data of LHCb with 9${\rm fb}^{-1}$ luminosity \cite{LHCb:2020gog} has provided the first measurement of the full set of CP-averaged angular observables in the isospin partner decay $B^+\to K^{+*}\mu^+\mu^-$. In this case the $K^{+*}$-meson reconstructed through the decay chain $K^{+*}\to K^0_S\pi^+$ with $K^0_S\to \pi^+\pi^-$. For a particular angular observable $P_2$ in the $6.0-8.0$ ${\rm GeV}^2$ interval there is the largest local disagreement with respect to the SM prediction. The corresponding deviation is around $3\sigma$. Considering this, using the definition given in \cite{Descotes-Genon:2013vna} we evaluate CP-averaged angular observables $P_2$ and $P'_5$. With these observables, we further impose the constraint in the parameter space of the MFS. We have considered the given CP-averaged binned data (for different $q^2$ values given in \cite{LHCb:2020gog} ) for the angular observables. The expressions for the observables $P_2$ and $P'_5$ containing NP contribution from the MFS can be written as,
\begin{eqnarray}
P_2&=&\beta_l\frac{I^s_6}{8I_2^s}~~~{\rm with}~~~I_6^s=2\beta_l\left[\re (\apaL^{}\apeL^*) - (L\to R) \right] \;,\\
\label{P2ex}
P_{5}^\prime&=&\frac{I_{5}}{2\sqrt{-I_2^c I_2^s}}~~~{\rm with}~~~I_5  = \sqrt{2}\beta_l\left[\re(\azeL^{}\apeL^*) - (L\to R)\right]\;,
\label{P5pex}
\end{eqnarray}
where, $I^{s}_2$ and $I^{c}_2$ are given in Eqs.~\ref{I2s} and \ref{I2c} respectively. With the given data and the corresponding theoretical predictions in MFS we have derived $\chi^2$ per degrees of freedom. Then using this we have imposed the condition on the parameter space which is allowed by 95\% C.L. of the given data for $P_2$ and $P'_5$. { Here, we would like to mention that in our analysis we have not considered the bin [6.0, 8.0] GeV$^2$ of $q^2$ since it is known to suffer from long distance $c\bar{c}$ corrections close to the open charm threshold.} 
\section{Numerical results}\label{neu_res}
\begin{figure}[t!]
\begin{center}
\hspace*{-0.5cm}\subfloat[]{\label{ab}\includegraphics[height=8cm,width=9.5cm,angle=0]{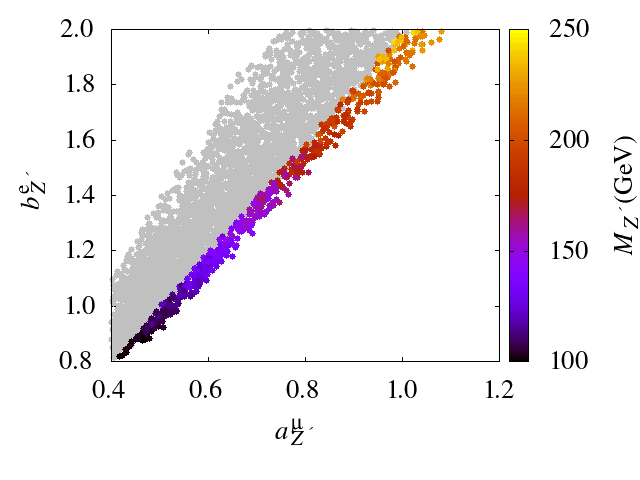}}
\subfloat[]{\label{MZpgbs}\includegraphics[height=8cm,width=9cm,angle=0]{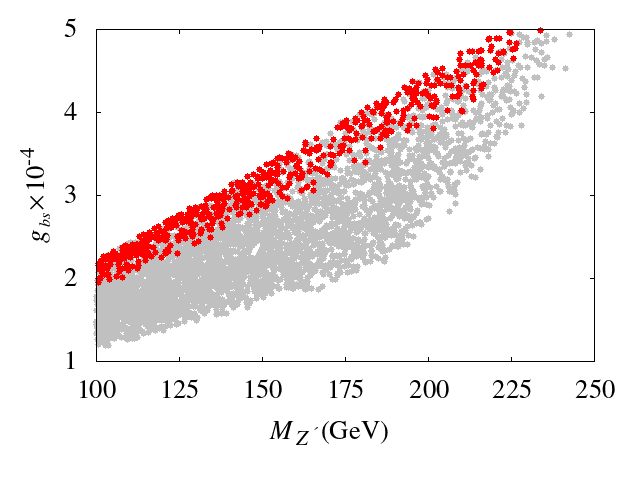}}
\caption{Left panel: Allowed parameter space in $a^\mu_{Z'}$ vs $b^e_{Z'}$ plane with allowed values of the mass $Z'$ boson (in GeV). Different values of masses are indicated by different colour code. Right panel: Allowed parameter space in $M_{Z'}$ (in GeV) vs $\textsl{g}_{bs}$ plane. For both the panel, we have taken the current experimental results of $(g-2)_\mu$, $(g-2)_e$ (with negative value of $\Delta a_e$), latest data of $R_K$, $R_{K^*}$, leading angular observables ($P_2$ and $P'_5$) of $B^+\to K^{+*}\mu^+\mu^-$ decay mode. Further, we have taken the constraint that is imposed by the ${B^0_s}-\bar{B^0_s}$ mixing. If we relax the angular observables conditions then there will be an enhancement of the allowed parameter space and which is reflected by the grey coloured region in both the panels.} 
\label{fig:all}
\end{center}
\end{figure} 

We have performed an extensive numerical scan sampling over $10^7$ points using the free parameters in MFS. From this scan, about four thousand ($0.04\%$) points survive which satisfy the latest measurement of $\Delta a_\mu$ at Fermilab \cite{Muong-2:2021ojo}, $\Delta a_e$ given by Barkeley laboratory \cite{Parker:2018vye}, $R_K$ published by LHCb collaboration \cite{LHCb:2021trn}, up to date results of $R_{K^{*}}$ for the both lower and central bin values of $q^2$ \cite{Aaij:2017vbb}. Moreover, we incorporate relevant constraint from ${B^0_s}-\bar{B^0_s}$ oscillation data \cite{Zyla:2020zbs}. Additionally we have also imposed the constraints from the leading angular observables ($P_2$ and $P'_5$) of $B^+\to K^{+*}\mu^+\mu^-$ decay mode \cite{LHCb:2020gog}.  In the left panel (\ref{ab}) of Fig.~\ref{fig:all} we have shown the allowed parameter space in $a^\mu_{Z'}$ vs $b^e_{Z'}$ plane with allowed values of the mass of $Z'$ boson (indicated by different colour code). This pattern can be explained if we consider the similar argument as given for the Fig.~\ref{fig:e_mu_a0e_b0mu}. However, in the case of Fig.~\ref{fig:all} apart from the case of $(g-2)_l$, we have to consider the NP contributions to the WCs $C_9$ and $C_{10}$ for $B$-meson decays. If one looks at the Eqs.~\ref{C9NP} and \ref{C10NP}, then it is evident that the NP contributions to WCs $C_9$ and $C_{10}$ are proportional to ${a^\mu_{Z'}}/{M^2_{Z'}}$ and ${b^e_{Z'}}/{M^2_{Z'}}$ respectively. Therefore, if the values of $a^\mu_{Z'}$ and $b^e_{Z'}$ are increased then in order to restrict the numerical prediction of the observables within the allowed range, the values of $M_{Z'}$ will also increase to suppress the propagator effect and it is depicted in the left panel (\ref{ab}) of Fig~\ref{fig:all}. Similar argument also holds good for the right panel (\ref{MZpgbs}) and it is evident from the Eqs.~\ref{C9NP} and \ref{C10NP}. Hence, following the previous argument and from the Eqs.~\ref{C9NP} and \ref{C10NP} it is clear that with the increasing values of $\textsl{g}_{bs}$ the values of $M_{Z'}$ will also increase and it is reflected from the right panel (\ref{MZpgbs}) of the Fig.~\ref{fig:all}. If we relax the constraints from angular observables of the $B^+\to K^{+*}\mu^+\mu^-$ decay mode, we expectedly obtain an enlarged allowed parameter space and these additional points are depicted in grey in both the panels of Fig.~\ref{fig:all}. 

\subsection{Some relevant constraints regarding $b\to s$ decay transitions}
Some comments on processes related $b\to s$ transitions are now in order. In our minimal scenario constraint from Br($B_s\to\mu^+\mu^-$) is not relevant, because it is dominated by the WC $C_{10}$. In this scenario, the WC $C_9$ gets the NP contribution for the $b\to s\mu^+\mu^-$ transition whereas the WC $C_{10}$ receives the NP contribution for $b\to s e^+e^-$ transition due to the presence of non zero value of $b^e_{Z'}$. Therefore, we have computed Br$(B_s\to e^+e^-)$ for allowed parameter points shown in the Fig.~\ref{fig:all}. Consequently, we have found that due to NP contribution there is a substantial amount of enhancement to the Br$(B_s\to e^+e^-)$ with respect to the corresponding SM prediction $8.6\times 10^{-14}$ \cite{LHCb:2020pcv}. This can be construed as an testable prediction of this framework. {For example, if we consider the value of $M_{Z'}$ is 180 GeV then within the allowed region of parameter space the model prediction for the Br$(B_s\to e^+e^-)$ can be as large as $2.43\times 10^{-12}$ which is well below the experimental upper limit value $9.4\times 10^{-9}$ \cite{LHCb:2020pcv}. Moreover, we have found that within the allowed values of $M_{Z'}$ (e.g., between 100 GeV to 200 GeV), the largest values of Br$(B_s\to e^+e^-)$ almost remain the same.}
Further, we have checked that parameter space (presented in the Fig. \ref{fig:all}) is also in consonance with the experimental results for the decays $B\to K^{(*)}e^+e^-$ \cite{LHCb:2013pra, LHCb:2014vgu}.

\section{Other experimental constraints}\label{excons}
We now turn our attention to other relevant bounds\footnote{In our analysis, we have not considered the constraints from flavour violating processes like $\mu\to e\gamma$ or $\tau\to 3\mu$ as there is no mixing in the charged lepton sector of the SM.} on the effective $Z'$ scenario from searches at both low energy and high energy collider experiments.

In most UV complete models where an exotic $Z'$ couples to the charged leptons, an interaction with the corresponding neutrinos is naturally expected. In fact in the limit of preserved ${\rm SU(2)}_L$ symmetry we expect identical coupling between the left handed charged leptons and their isospin partner neutrinos.  Additional constraints  that  arise due to the coupling of the $Z'$ to the neutrinos are summarised below:
 
\begin{enumerate}
\item The CCFR experiment has put stringent bounds on muon neutrino-nucleus  scattering cross section ($\nu_{\mu} (\overline{\nu_{\mu}}) + N \rightarrow \nu_{\mu} (\overline{\nu_{\mu}}) + \mu^+ \mu^- + N$) that can provide constraint on the parameter space of interest. The neutrino trident production cross section measured at the CCFR experiment at  ${\sigma_{\rm CCFR}}/{\sigma_{\rm SM}} = 0.82\pm 0.28$ \cite{Mishra:1991bv}. 

\item Measurement of Br($B\to K^{(*)}\nu\bar{\nu}$) by the Belle collaboration \cite{Belle:2017oht} can also potentially constrain the $Z'-\nu$ couplings. 

\end{enumerate}
It is known that the neutrino trident production cross section excludes the $(g-2)_\mu$ allowed region above the GeV scale for ${\rm SU(2)}_L$ invariant couplings \cite{Altmannshofer:2014pba, Altmannshofer:2019zhy}. Continuing with our bottom up approach we introduce  a generic coupling of the $Z'$ with the neutrinos aligned with the charged lepton coupling introduced in Sec.~\ref{model}. We keep the couplings of the $Z'$ to the charged leptons and the corresponding neutrinos independent and their ratio will be parametrised by $a_{Z'}^{\nu_l} / a_{Z'}^l   =\varepsilon_l$. The $\varepsilon_l$ is a measure of the  isospin violation in the couplings  and  $\varepsilon_l=1$ represents the ${\rm SU(2)}_L$ invariant limit. In the passing, we note that UV complete models with ${\rm SU(2)}_L$ violating $Z'$ couplings are not very common. It is possible to construct scenarios where the ${\rm SU(2)}_L$ violating couplings of SM leptons to the $Z'$ entirely originate from a linear mixing with exotic vector like lepton partners. Provided the charged lepton partner has a different ${\rm U(1)}'$ quantum number compared to the corresponding neutrino partner, the effective $Z'$ couplings with the SM leptons will violate isospin after electroweak symmetry breaking. These frameworks can possibly be embedded in UV complete scenarios. For example see ref.\;\cite{Li:2019sty} for an $E_6$ GUT scenario where several exotic scalar fields, that obtain vacuum expectation values as $E_6$ breaks to the SM gauge group, drive a linear mixing between the SM matter fields with their vector like partners. While the focus in \cite{Li:2019sty} is on the isospin violating coupling in the quark sector, a generalisation to the leptonic sector with a Dirac like neutrino mass is straightforward. For other approach to isospin violation in the quark sector see for example the discussion in the context of dark matter phenomenology in ref.\;\cite{Frandsen:2011cg, Li:2022qrl}, which can potentially be extended to the leptonic sector.

Following \cite{Altmannshofer:2014rta} we evaluate the neutrino trident production cross section assuming $\varepsilon_\mu(\neq 1)$ and compare the parameter space of interest with the CCFR results.  Further, we utilise {\tt flavio} \cite{Straub:2018kue} to numerically evaluate Br($B\to K^{(*)}\nu\bar{\nu}$) \cite{Buras:2014fpa} in the parameter space that is consistent with $(g-2)_\mu$, $(g-2)_e$, $R_{K^{(*)}}$, leading angular observables of the decay $B^+\to K^{+*}\mu^+\mu^-$, ${B^0_s}-\bar{B^0_s}$ mixing and CCFR data\footnote{An additional constraint from the $\nu_e e$ scattering \cite{Bellini:2011rx, Borexino:2013zhu, Borexino:2017rsf} can also play a role for MeV scale $Z'$ model. However, the effect for $Z'$ masses greater than 1.5 GeV is negligible \cite{Lindner:2018kjo}. As we concentrate on the weak scale $Z'$ that is relevant for the $B$-meson sector we do not consider this limit in our analysis.}. 

\begin{figure}[t!]
\begin{center}
\includegraphics[height=9cm,width=12cm,angle=0]{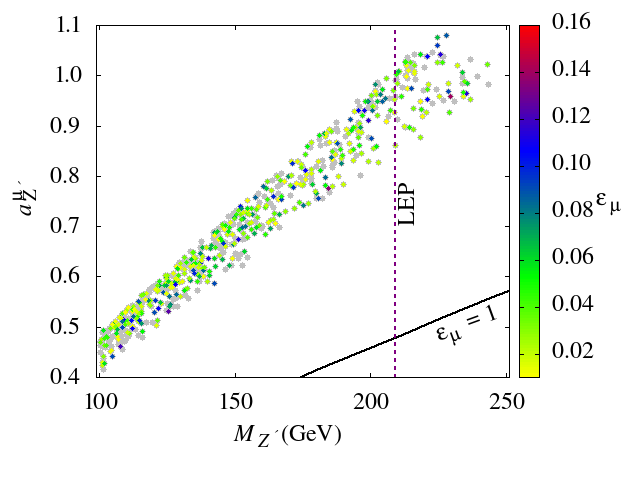}
\caption{Allowed parameter space in $M_{Z'}-a^\mu_{Z'}$ plane with allowed values of the isospin violating parameter $\varepsilon_\mu$. Different values of $\varepsilon_\mu$ are indicated by different colour codes. The parameter space has been obtained by considering the current experimental results of $(g-2)_\mu$, $(g-2)_e$ (with negative value of $\Delta a_e$), latest data of $R_K$, $R_{K^*}$, leading angular observables ($P_2$ and $P'_5$) of $B^+\to K^{+*}\mu^+\mu^-$ decay mode and ${B^0_s}-\bar{B^0_s}$ oscillation data. Further we imposed the CCFR data for neutrino trident production. Within the allowed parameter points the branching ratios for $B\to K^{(*)}\nu\bar{\nu}$ is found to be consistent with the experimental upper bound $1.6 (2.7)\times 10^{-5}$ and remain  one order of magnitude below this limit in the entire region of the parameter space of interest. If we relax the constraint from neutrino trident production cross section then there will be an enhancement of the allowed parameter space and which is reflected by the grey coloured region. The purple coloured vertical line represents the LEP collider bound on $M_{Z'}$.} 
\label{fig:CCFR}
\end{center}
\end{figure}

In Fig.~\ref{fig:CCFR} we present the parameter space which is allowed by  experimental data considered in Sec.~\ref{neu_res} and additionally is in agreement with the CCFR data for neutrino trident production and the limits on  Br($B\to K^{(*)}\nu\bar{\nu}$). Expectedly the ${\rm SU(2)}_L$ invariant couplings are excluded by the CCFR data. This necessitates the introduction of non-trivial isospin violation in the $Z'$ couplings parametrised by $\varepsilon_l \neq 1$. In Fig.~\ref{fig:CCFR} the allowed parameter points in the $M_{Z'}-a^\mu_{Z'}$ plane  have values of $\varepsilon_\mu$ represented by different colours. The black line represents the CCFR exclusion limit for  $\varepsilon_\mu=1$. As can be read off from the plot the region of parameter space consistent with $(g-2)_\mu, B$-physics  observable and CCFR require $\varepsilon_\mu < 0.2.$ For the allowed parameter points the branching ratios for $B\to K^{(*)}\nu\bar{\nu}$ is found to be consistent with the experimental upper bound $1.6 (2.7)\times 10^{-5}$ \cite{Belle:2017oht} and remain  one order of magnitude below this limit in the entire region of the parameter space of interest.

A few comments about the direct collider bounds on the $Z'$ model considered here is now in order. The most stringent collider bound on the effective framework arises from the LHC searched in the $pp\to Z\to 4\mu$  channel and is relevant in the range $5\lesssim M_{Z'}\lesssim 70\,\mathrm{GeV}$ \cite{Falkowski:2018dsl,Altmannshofer:2014cfa,Altmannshofer:2014pba,Altmannshofer:2016jzy} which is not of concern for the parameter space presented in Fig.~\ref{fig:CCFR}.  Given the coupling between the electron and $Z'$ in the MFS the most relevant constraint from LEP  \cite{Zyla:2020zbs} (indicated by purple coloured vertical dashed line in Fig.~\ref{fig:CCFR}) exclude the parameter space below $M_{Z'}< 209$ GeV.  The Fig.~\ref{fig:CCFR} clearly indicates that some of the sampled parameter points are able to survive all the constraints considered in this study provided an isospin violating coupling is assumed between the exotic $Z'$ and the lepton doublets.

\section{LKB measurement of $(g-2)_{e}$}\label{plus_g2e}
Before we conclude we would like to remark on a recent measurement at Laboratoire Kastler Brossel
(LKB) with rubidium atoms reported a new value for the fine structure constant \cite{Morel:2020dww}. Using this measurement, the SM prediction of $(g-2)_e$ shift, and is estimated to be $1.6\sigma$ lower with respect to the experimental value \cite{Hanneke:2008tm} with,
\begin{equation}
\Delta a_e = a_e^{\rm exp} - a_e^{\rm SM} = (4.8\pm 3.0)\times 10^{-13}.
\label{eg2positive}
\end{equation}

A discussion about the minimal flavoured $Z'$ scenario in view of this recent result is now in order.

\begin{figure}[htbp!]
\begin{center}
\subfloat[]{\label{fig:aZMZ_pluse}\includegraphics[height=7.5cm,width=9cm,angle=0]{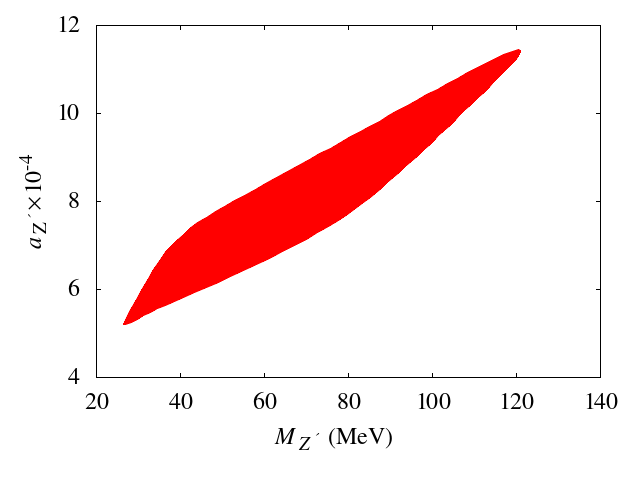}}
\subfloat[]{\label{fig:aZbZ_plus_g2}\includegraphics[height=7.5cm,width=9.9cm,angle=0]{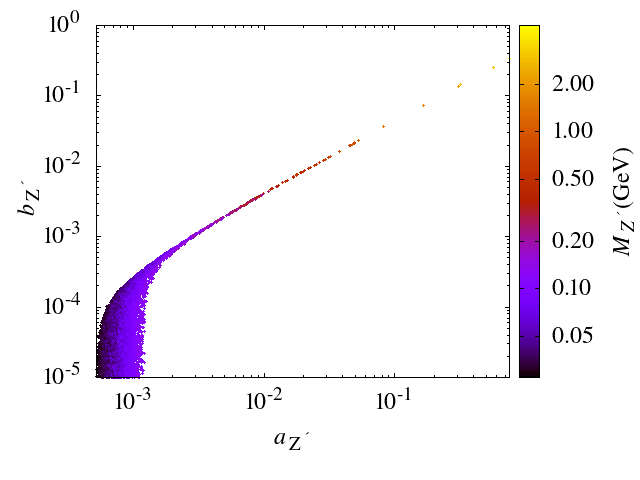}}\\
\subfloat[]{\label{fig:aZeaZmu_plus_g2}\includegraphics[height=7.5cm,width=9.8cm,angle=0]{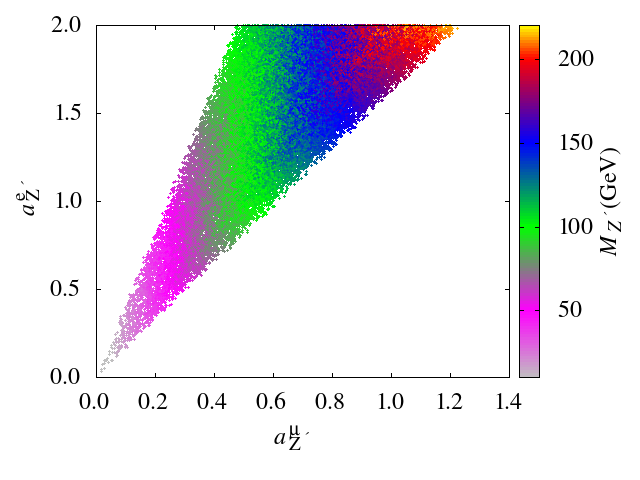}}
\hspace*{-0.2cm}\subfloat[]{\label{fig:aZeaZmuMZpgbs_plus_g2}\includegraphics[height=7.5cm,width=9cm,angle=0]{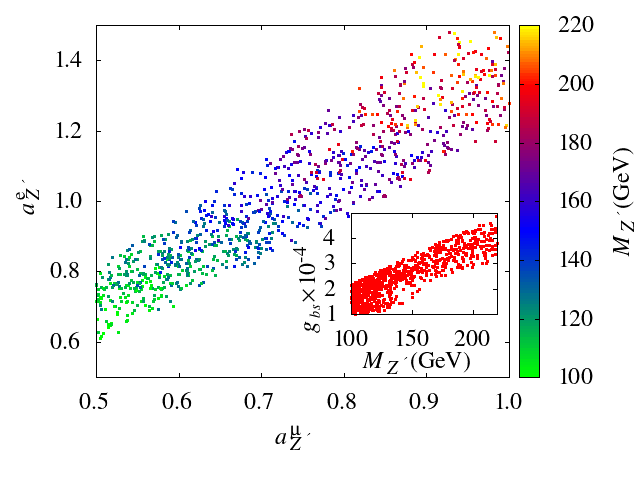}}
\caption{Upper left panel: 1$\sigma$ allowed parameter space in $M_{Z^{\prime}}$ (in MeV) vs $a_{Z^{\prime}}$ plane satisfied by both the data of $(g-2)_{\mu}$ and $(g-2)_{e}$. Upper right panel: 1$\sigma$ allowed parameter space in $a_{Z^{\prime}}$ vs $b_{Z^{\prime}}$ (same for muon and electron) plane with the allowed values of $M_{Z^{\prime}}$ (in GeV) satisfied by both the data of $(g-2)_{\mu}$ and $(g-2)_{e}$. Lower left panel: 1$\sigma$ allowed parameter space in $a^\mu_{Z^{\prime}}$ vs $a^e_{Z^{\prime}}$ plane with allowed values of $M_{Z^{\prime}}$ (in GeV). Lower right panel: parameter space in $a^\mu_{Z^{\prime}}$ vs $a^e_{Z^{\prime}}$ plane with allowed values of $M_{Z^{\prime}}$ (in GeV) allowed by $\Delta a_\mu$ by Fermilab, result on $\Delta a_e$ given by LBL, latest data of $R_K$ published by LHCb collaboration, up to date values of $R_{K^{*}}$ for the both lower and central bin values of $q^2$ and leading angular observables of the decay mode $B^+\to K^{+*}\mu^+\mu^-$. The inset shows the region in $M_{Z^{\prime}}$ (in GeV) vs $\textsl{g}_{bs}$ plane allowed by the above mentioned experimental results.}
\label{psoitive_I_II}
\end{center}
\end{figure}

\begin{enumerate}
\item With both positive value of $\Delta a_e$ and $\Delta a_\mu$ one can hope to explain both simultaneously in a scenario where both the electron and muon have identical vectorial coupling with the $Z'$ as defined in Eq.~\ref{aZp_mu1}. The corresponding $1\sigma$ allowed parameter space is shown in $M_{Z^{\prime}}$ vs $a_{Z^{\prime}}$ plane in the Fig.~\ref{fig:aZMZ_pluse}. The preferred  $M_{Z^{\prime}}$ in the MeV scale is too restricted to explain the LFUV in the $B$-meson sector.

\item A possibility is where both the electron and muon have vectorial as well as axial-vectorial coupling (of same strengths) with the $Z'$. In such case the most recent data of $(g-2)_e$ with positive value of $\Delta a_e$ and the recent data of $(g-2)_{\mu}$ can be explained simultaneously. In Fig.~\ref{fig:aZbZ_plus_g2}, $1\sigma$ allowed parameter space has been shown in $a_{Z^{\prime}}$ vs $b_{Z^{\prime}}$ plane for allowed values of $M_{Z^{\prime}}$. Again in this case also, the allowed values of independent parameters are restricted within very small values.

\item In the passing we note that the other possible four parameter model is where, $a^\mu_{Z^{\prime}}\neq 0$, $b^\mu_{Z^{\prime}}= 0$ (for muon) but $a^e_{Z^{\prime}}= 0$, $b^e_{Z^{\prime}}\neq 0$ (for electron), then it is not possible to explain both the $(g-2)_{\mu}$ and $(g-2)_e$ with positive value of $\Delta a_e$ simultaneously. Because, in such scenario $(g-2)_e$ with positive value of $\Delta a_e$ cannot be explained.

\item We further consider a four parameter scenario in which both the electron and muon have independent vectorial coupling with the $Z'$. In Fig.~\ref{fig:aZeaZmu_plus_g2} we have shown the 1$\sigma$ allowed parameter space satisfied by both the $(g-2)_\mu$ and $(g-2)_e$ with positive value of $\Delta a_e$. From this figure it is clear that the mass of the $Z'$ can be increased substantially with respect to the previous scenarios making it more favourable to explain LFUV in the $B$-sector. We compute the LFUV observables $R_{K^{(*)}}$, ${B^0_s}-\bar{B^0_s}$ mass difference and leading angular observables of the decay mode $B^+\to K^{+*}\mu^+\mu^-$ with the motivation to find out the region of parameter space which satisfy the corresponding experimental results simultaneously. The result from our analysis is depicted in the Fig.~\ref{fig:aZeaZmuMZpgbs_plus_g2}. The inset shows the corresponding allowed region in $M_{Z^{\prime}}$ (in GeV) vs $\textsl{g}_{bs}$ plane. 

We expectedly find that the identification of the most optimistic flavoured $Z'$ model depends on the relative sign of the $\Delta a_\mu$ and $\Delta a_e$.

\end{enumerate}
{The numerical results presented here from our in-house implementation of the $B$-physics observables have been extensively validated with the results obtained from the publicly available package {\tt flavio} \cite{Straub:2018kue}. We reproduce Fig.~\ref{ab}, Fig.~\ref{MZpgbs}, Fig.~\ref{fig:CCFR} and Fig.~\ref{fig:aZeaZmuMZpgbs_plus_g2} using the package {\tt flavio}. A detailed quantitative comparison of our results with {\tt flavio} is presented in Appendix \ref{flavio}.}
\section{Conclusion}\label{concl}
A synergy of experimental results in measurement of $R_K$ (with $3.1\sigma$ deviation) and $R_{K^{*}}$ by LHCb collaboration, $(g-2)_\mu$ (with $4.2\sigma$ deviation) by Fermi Lab and $(g-2)_e$ provide a tantalizing hint of lepton flavour violation and hence Beyond Standard Model Physics in the flavour sector.

{In this paper, instead of conforming to a specific UV complete scenario we survey the data driven phenomenological effective models with vectorial and axial-vectorial leptonic coupling for the $Z'$}. We systematically identify the minimal flavoured $Z'$ model that can simultaneously explain these experimental evidences of lepton flavour universality violation while remaining in consonance with the correlated ${B^0_s}-\bar{B^0_s}$ oscillation. We explore the parameter space that is allowed by these observables taking into account the leading angular observables of the decay mode $B^+\to K^{+*}\mu^+\mu^-$.

From our systematic study we observe that the models are very sensitive to the relative sign between $\Delta a_\mu$ and $\Delta a_e$. For example, we find that a $Z'$ that couple vectorially to moun while having an purely axial-vectorial coupling to electron can explain the data of anomalous magnetic moment of leptons (muon and electron). An off diagonal coupling to the quarks can simultaneously explain the $B$-physics observables for a weak scale $Z'$. An increase in the resolution of measurement of the anomalous magnetic moment of lepton in the future will provide a handle in identifying specific scenarios of flavoured $Z'$ models. On the other hand we find the minimal model with flavour specific vectorial coupling to the lepton suits the measurement of $\Delta a_e$ using the LKB data.

Interestingly the  CCFR data for neutrino trident production  cross section excludes an ${\rm SU(2)}_L$ invariant coupling between the exotic  $Z'$ and the leptonic doublets for models that simultaneously satisfy the $B$-physics and $(g-2)_l$ constraints. This implies the uncomfortable reality of  an isospin violating couplings for the $Z'$ along with flavour violation.

{\bf Acknowledgements} We would like to give thank Chirashree Lahiri and Rohan Pramanick for computational and technical support.  AS acknowledges the financial support from Department of Science and Technology, Government of India through SERB-NPDF scholarship with grant no.:PDF/2020/000245. TSR acknowledges Department of Science and Technology, Government of India, for support under grant agreement no.:ECR/2018/002192 [Early Career Research Award].\\


{\Large{\bf Appendix}}

\begin{appendix}
\renewcommand{\thesection}{\Alph{section}}
\renewcommand{\theequation}{\thesection-\arabic{equation}} 

\setcounter{equation}{0} 

\section{Relevant Wilson Coefficients for the \boldmath$b\to s l^+l^-$ transitions}\label{NDR}
In this appendix we collect all the relevant Wilson Coefficients that are useful in constructing the observables related to the $b\to s l^+l^-$ transitions. The operator $\mathcal{O}_{10}$ does not evolve under QCD renormalisation and  its coefficient is independent of energy scale $Q$ and can be expressed in the following way
\begin{equation}
C_{10}(Q) = - \frac{Y(x_t )}{\sin^2\theta_{W}}+C^{\rm NP}_{10}\;,
\end{equation}
where $\theta_W$ is the Weinberg angle.  Unlike $C_{10}$, $C_9$ varies with energy scale and using the results of NLO QCD corrections to $C^{\rm eff}_{9}(Q)$ in the SM \cite{Misiak:1992bc, Buras:1994dj} we can readily obtain this coefficient in the NP scenario under the naive dimensional regularisation (NDR) renormalisation scheme as
\begin{eqnarray}\label{c9_eff}
C_9^{\rm eff}(q^2)&=&C_9^{\rm NDR}\tilde{\eta}\left(\frac{q^2}{m^2_b}\right)+h\left(z,\frac{q^2}{m^2_b}\right)\left(3C_1+C_2+3C_3+C_4+3C_5+C_6\right) \\ \nonumber 
&&-\frac 12 h\left(1,\frac{q^2}{m^2_b}\right)\left(4C_3+4C_4+3C_5+C_6\right)-\frac 12 h\left(0,\frac{q^2}{m^2_b}\right)\left(C_3+4C_4\right)  \\ \nonumber 
&&+\frac 29\left(3C_3+C_4+3C_5+C_6\right),
\end{eqnarray}
where,
\begin{equation}\label{C9tilde}
C_9^{\rm NDR}  =  
P_0^{\rm NDR} + \frac{Y(x_t )}{\sin^2\theta_{W}}+C^{\rm NP}_9 -4 Z(x_t ) + P_E E(x_t )\;.
\end{equation}
The value\footnote{The analytic formula for $P_0^{\rm NDR}$ has been given in
\cite{Buras:1994dj}.} of $P_0^{\rm NDR} (P_E)$ is set at $2.60\pm 0.25$ \cite{Buras:2003mk} (${\cal O}(10^{-2})$ \cite{Buras:1994dj}). The function $Y(x_t)$, $Z(x_t)$ and $E(x_t)$ are the usual Inami-Lim functions \cite{Buras:1994dj, Buchalla:1995vs}.
The function $\tilde{\eta}$ (given in the Eq.\;\ref{c9_eff}) represents single gluon corrections to the matrix element $\mathcal{O}_{9}$ and it takes the form \cite{Buras:1994dj} 
\begin{eqnarray}
\tilde{\eta}\left(\frac{q^2}{m^2_b}\right)=1+\frac{\alpha_s}{\pi}\omega\left(\frac{q^2}{m^2_b}\right),
\end{eqnarray}
where $\alpha_s$ is the QCD fine structure constant. The functional forms of $\omega$ and $h$ are given by \cite{Buras:1994dj}
\begin{eqnarray}
\omega\left(\frac{q^2}{m^2_b}\right)&=&-{2\over9}\pi^2-{4\over3}{\rm Li}_{_2}\left(\frac{q^2}{m^2_b}\right)-{2\over3}\ln \left(\frac{q^2}{m^2_b}\right)\ln\left(1-\frac{q^2}{m^2_b}\right) \\ \nonumber 
&&-{5+4\frac{q^2}{m^2_b}\over3\bigg(1+2\frac{q^2}{m^2_b}\bigg)}\ln\left(1-\frac{q^2}{m^2_b}\right) -{2\frac{q^2}{m^2_b}\bigg(1+\frac{q^2}{m^2_b}\bigg)\bigg(1-2\frac{q^2}{m^2_b}\bigg)\over3\bigg(1-\frac{q^2}{m^2_b}\bigg)^2\bigg(1+2\frac{q^2}{m^2_b}\bigg)}\ln \left(\frac{q^2}{m^2_b}\right) \\ \nonumber 
&&+{5+9\frac{q^2}{m^2_b}-6\left(\frac{q^2}{m^2_b}\right)^2\over6\bigg(1-\frac{q^2}{m^2_b}\bigg)\bigg(1+2\frac{q^2}{m^2_b}\bigg)}\;,
\end{eqnarray}
and
\begin{eqnarray}
\label{hz}
h\left(z,\frac{q^2}{m^2_b}\right)&=&{8\over27}-{8\over9}\ln{m_{_b}\over\mu}-{8\over9}\ln z+{16z^2m^2_b\over9q^2} \\ \nonumber 
&&-{4\over9}\left(1+{2z^2m^2_b\over q^2}\right)\sqrt{\bigg|1-{4z^2m^2_b\over q^2}\bigg|}
\left\{\begin{array}{ll}\ln\Bigg|{\sqrt{1-\frac{4z^2m^2_b}{q^2}}+1\over\sqrt{1-\frac{4z^2m^2_b}{q^2}}-1}\Bigg|-i\pi,
&{\rm if}\: \frac{4z^2m^2_b}{q^2}<1\\
2\arctan{1\over\sqrt{\frac{4z^2m^2_b}{q^2}-1}},&{\rm if}\: \frac{4z^2m^2_b}{q^2}>1\end{array}\right.\;. 
\end{eqnarray}
Wilson Coefficients {$C_1\ldots C_6$} are defined as \cite{Buras:1994qa}
\begin{eqnarray}
\label{c1}
C_1(M_W)&=&\frac{11}{2}\frac{\alpha_s(M_W)}{4\pi}\;, \\
\label{c2}
C_2(M_W)&=&1-\frac{11}{6}\frac{\alpha_s(M_W)}{4\pi}\;,\\
\label{c3_4}
C_3(M_W)&=&-\frac 13 C_4(M_W)=-\frac{\alpha_s(M_W)}{24\pi}=\widetilde{E}(x_t )=E(x_t )-\frac 23\;,\\
C_5(M_W)&=&-\frac 13 C_6(M_W)=-\frac{\alpha_s(M_W)}{24\pi}=\widetilde{E}(x_t )=E(x_t )-\frac 23\;.
\label{c5_5}
\end{eqnarray}
The formula of decay branching ratio of $B\to K^{(*)} l^+l^-$ consists of another effective Wilson Coefficient namely $C_{7}^{{\rm eff}}$ for which there is no NP contribution in our chosen scenario. Within the SM $C_{7}^{{\rm eff}}$ can be defined as \cite{Buras:1994dj}
\begin{eqnarray}
\label{C7eff}
C_{7}^{{\rm eff}} & = & 
-\frac{1}{2} \eta^\frac{16}{23} D'(x_t )-\frac{1}{2} \frac{8}{3}
\left(\eta^\frac{14}{23} - \eta^\frac{16}{23}\right) E'(x_t ) +
 C_2(M_W)\sum_{i=1}^8 h_i \eta^{a_i},
\end{eqnarray}
with
\begin{equation}
\eta  =  \frac{\alpha_s(M_W)}{\alpha_s(m_b)},~~~\alpha_s(m_b) = \frac{\alpha_s(M_Z)}{1 
- \frac{23}{3} \frac{\alpha_s(M_Z)}{2\pi} \, \ln(M_Z/m_b)}. 
\label{eq:asmumz}
\end{equation}
The values of $a_i$, $h_i$ and $\bar h_i$ can be obtained from \cite{Buras:1994dj}. $D'(x_t )$ and $E'(x_t )$ are the Inami-Lim functions \cite{Buras:1994dj, Buchalla:1995vs} that represent SM contributions (at the LO level) to the photonic and gluonic magnetic dipole moment operators.


\section{Form Factor for the \boldmath$B\to K^{(*)} l^+l^-$ transitions}
In this appendix we briefly summarise the $B\to K^{(*)}$ form factors related to the rare $B$-meson decays considered in our analysis. 

\subsection{\boldmath Details of form factors for $B^+\to K^+l^+l^-$ transitions}
\label{sec:ftfprel}
The long-distance effects for hadronic dynamics of $B^+\to K^+l^+l^-$ decay is represented by the following matrix elements \cite{Bartsch:2009qp}
\begin{eqnarray}
\langle K^+(p')|\bar s\gamma^\mu b|B^+(p)\rangle
&=& f_+(s)\, (p+p')^\mu +[f_0(s)-f_+(s)]\,\frac{m^2_B-m^2_K}{q^2}q^\mu\,,
\label{fpf0def}\\
\langle K^+(p')|\bar s\sigma^{\mu\nu}b|B^+(p)\rangle
&=& i\frac{f_T(s)}{m_B+m_K}\left[(p+p')^\mu q^\nu - q^\mu (p+p')^\nu\right]\,.
\label{ftdef}
\end{eqnarray}
Here, the form factors are $f_+$, $f_0$ and $f_T$. Further $q=p-p'$ and $s=q^2/m^2_B$. $f_0$ terms drops out from the expression of differential decay width (see Eq.~\ref{btokdw}) due to smallness of lepton masses. Using the approach given in ref.~\cite{Buras:2014fpa} we implement the following expression for $f_+$,
\begin{equation}
f_+(q^2) = \frac{1}{1-q^2/m_+^2}\left[
\alpha_0 + \alpha_1 z(q^2) + \alpha_2 z^2(q^2)+\frac{z^3(q^2)}{3}(-\alpha_1+2\alpha_2)
\right],
\end{equation}
with a simplified series expansion (SSE)
\begin{equation}
z(t) = \frac{\sqrt{t_+-t}-\sqrt{t_+-t_0}}{\sqrt{t_+-t}+\sqrt{t_+-t_0}}\,,
\end{equation}
where, $t_\pm=(m_B\pm m_K)^2$ and $t_0=t_+(1-\sqrt{1-t_-/t_+})$.
The resonance mass is given by $m_+=m_B+0.046$\,GeV. The values of the parameters $\alpha_0$, $\alpha_1$, and $\alpha_2$ as are given bellow \cite{Buras:2014fpa}
\begin{align}
\alpha_0 &= 0.432 \pm 0.011
\,,&
\alpha_1 &= -0.664 \pm 0.096
\,,&
\alpha_2 &= -1.20 \pm 0.69
\,.
\end{align}
The corresponding expression for $f_T$ is extracted from the following ratio,
\begin{equation}\label{ftfp}
\frac{f_T(s)}{f_+(s)}=\frac{m_B+m_K}{m_B}\,.
\end{equation}
This is independent of unknown hadronic quantities in the domain of interest \cite{Charles:1998dr,Beneke:2000wa, Wise:1992hn,Burdman:1992gh,Falk:1993fr,Casalbuoni:1996pg,Buchalla:1998mt}.


\subsection{Details of form factors for $B^0\to K^{0*}l^+l^-$ transitions}\label{BSZ:ff}

The matrix elements for the relevant operators for $B^0(p)\to K^{0*}(k)$ transitions in
terms of momentum transfer ($q^\mu = p^\mu - k^\mu$) dependent form factors can be written as \cite{Altmannshofer:2008dz}
\begin{eqnarray}
\lefteqn{
\langle K^{0*}(k) | \bar s\gamma_\mu(1-\gamma_5) b | B^0(p)\rangle  =  
-i \epsilon^*_\mu (m_B+m_{K^*})
A_1(q^2) + i (2p-q)_\mu (\epsilon^* \cdot q)\,
\frac{A_2(q^2)}{m_B+m_{K^*}}}\hspace*{2.8cm}\nonumber\\
&& {}+  i
q_\mu (\epsilon^* \cdot q) \,\frac{2m_{K^*}}{q^2}\,
\left[A_3(q^2)-A_0(q^2)\right] +
\epsilon_{\mu\nu\rho\sigma}\epsilon^{*\nu} p^\rho k^\sigma\,
\frac{2V(q^2)}{m_B+m_{K^*}},\hspace*{0.5cm}\label{eq:SLFF}
\end{eqnarray}
and,
\begin{eqnarray}
\lefteqn{\langle K^{0*}(k) | \bar s \sigma_{\mu\nu} q^\nu (1+\gamma_5) b |
B^0(p)\rangle = i\epsilon_{\mu\nu\rho\sigma} \epsilon^{*\nu}
p^\rho k^\sigma \, 2 T_1(q^2)}\nonumber\\
& {} + T_2(q^2) \left[ \epsilon^*_\mu
  (m_B^2-m_{K^*}^2) - (\epsilon^* \cdot q) \,(2p-q)_\mu \right] + T_3(q^2) 
(\epsilon^* \cdot q) \left[ q_\mu - \frac{q^2}{m_B^2-m_{K^*}^2}(2p-q)_\mu
\right].\nonumber\\[-10pt]\label{eq:pengFF}
\end{eqnarray}
Here, $\epsilon_\mu$ represents polarization vector of
the $K^*$. The form factors $A_i$ and $V$ are scale independent. On the other hand the $T_i$ depend on the renormalisation scale. 
The form factor in the light cone sum rules (LCSR) scheme can be generically written as \cite{Bharucha:2015bzk}
\begin{equation}
F_i(q^2) = P_i(q^2) \sum_k \alpha_k^i \,\left[z^*(q^2)-z^*(0)\right]^k\,,
\label{eq:SSE}
\end{equation}
with an SSE, 
\begin{equation}
z^*(t) = \frac{\sqrt{t_+-t}-\sqrt{t_+-t_0}}{\sqrt{t_+-t}+\sqrt{t_+-t_0}} \;,
\end{equation}
where,
$t_\pm \equiv (m_B\pm m_{K^*})^2$ and $t_0\equiv t_+(1-\sqrt{1-t_-/t_+})$.
Here, $P_i(q^2)=(1-q^2/m_{R,i}^2)^{-1}$ represents a simple pole corresponding to the first resonance 
in the spectrum. Appropriate resonance masses $m_{R,i}$ and the coefficients $\alpha^i_k$ can be extracted from \cite{Bharucha:2015bzk}.

\section{Transversity amplitudes}
\label{trvnampli}
The expressions of the transversity amplitudes (up to corrections of $\mathcal{O}(\alpha_s)$) in terms of appropriate Wilson Coefficients and form factors are given as follows \cite{Altmannshofer:2008dz}

\begin{equation}
A_{\perp L,R}  =  N \sqrt{2} \lambda^{1/2} \bigg[ 
\left[ C_9^\eff \mp C_{10} \right] \frac{ V(q^2) }{ m_B + m_\kstar} 
 + \frac{2m_b}{q^2} C_7^\eff T_1(q^2)
\bigg], \label{Applr}
\end{equation}
\begin{equation}
A_{\parallel L,R}   = - N \sqrt{2}(m_B^2 - m_\kstar^2) \bigg[ \left[ C_9^\eff \mp C_{10} \right] 
\frac{A_1(q^2)}{m_B-m_\kstar}
+\frac{2 m_b}{q^2} C_7^\eff T_2(q^2)
\bigg], \label{Aprlr}
\end{equation}
\bea
A_{0L,R}  =  - \frac{N}{2 m_\kstar \sqrt{q^2}}  \bigg\{ 
 \left[ C_9^\eff \mp C_{10} \right]
\bigg[ (m_B^2 - m_\kstar^2 - q^2) ( m_B + m_\kstar) A_1(q^2) 
 -\lambda \frac{A_2(q^2)}{m_B + m_\kstar}
\bigg] 
\nonumber\\
 + {2 m_b}C_7^\eff \bigg[
 (m_B^2 + 3 m_\kstar^2 - q^2) T_2(q^2)
-\frac{\lambda}{m_B^2 - m_\kstar^2} T_3(q^2) \bigg]
\bigg\},\label{A0lr}
\eea
\begin{equation}
 A_t  = \frac{N}{\sqrt{q^2}}\lambda^{1/2}2 C_{10}A_0(q^2),
\label{At}
\end{equation}

with 
\begin{equation}
N= V_{tb}^{\vphantom{*}}V_{ts}^* \left[\frac{G_F^2 \alpha^2_{\rm em}}{3\cdot 2^{10}\pi^5 m_B^3}
 q^2 \lambda^{1/2}
\beta_l \right]^{1/2},
\end{equation}
where $\lambda= m_B^4  + m_{K^*}^4 + q^4 - 2 (m_B^2 m_{K^*}^2+ m_{K^*}^2 q^2  + m_B^2 q^2)$ and $\beta_l=\sqrt{1-4m_l^2/q^2}$. Moreover, $L$ and $R$ refer to the chirality of the leptonic current. Here the particular amplitude $A_t$ is related to the time-like component of the virtual $K^*$, and it does not contribute in the case of massless leptons. Therefore, it can be neglected if the lepton mass is small in comparison to the mass of the lepton pair. 

\vspace*{-0.5cm}

{
\section{Relative comparison with flavio}\label{flavio}
In this appendix we present a detailed comparison of numerical results obtained from our in-house implementation and the publicly available package {\tt flavio} \cite{Straub:2018kue}. 
\begin{figure}[htbp!]
\begin{center}
\hspace*{-0.5cm}\subfloat[]{\label{ab_flavio}\includegraphics[height=7.5cm,width=8.5cm,angle=0]{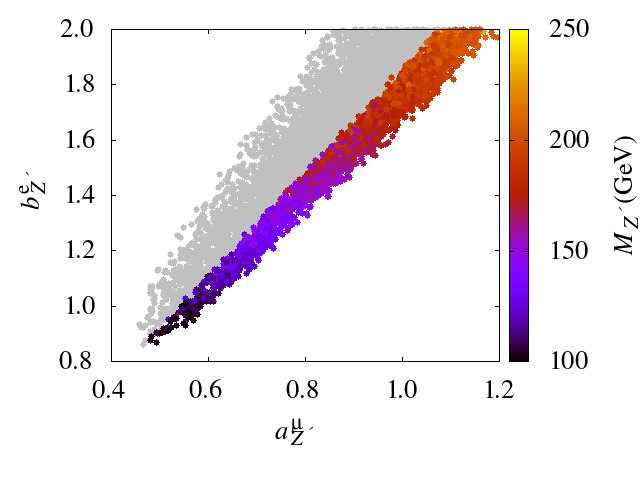}}
\subfloat[]{\label{MZpgbs_flavio}\includegraphics[height=7.5cm,width=8cm,angle=0]{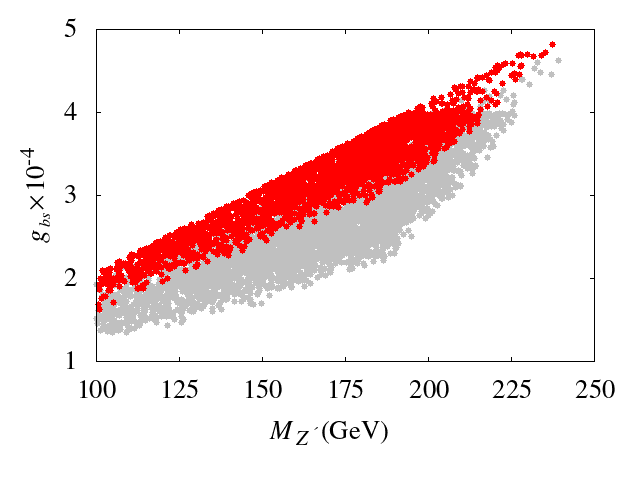}}\\
\subfloat[]{\label{fig:flavio_CCFR}\includegraphics[height=7.5cm,width=8cm,angle=0]{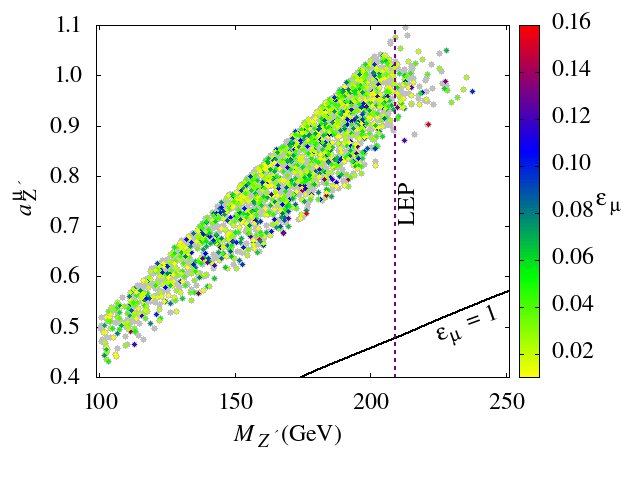}}
\subfloat[]{\label{fig:flavio_pos}\includegraphics[height=7.5cm,width=8.5cm,angle=0]{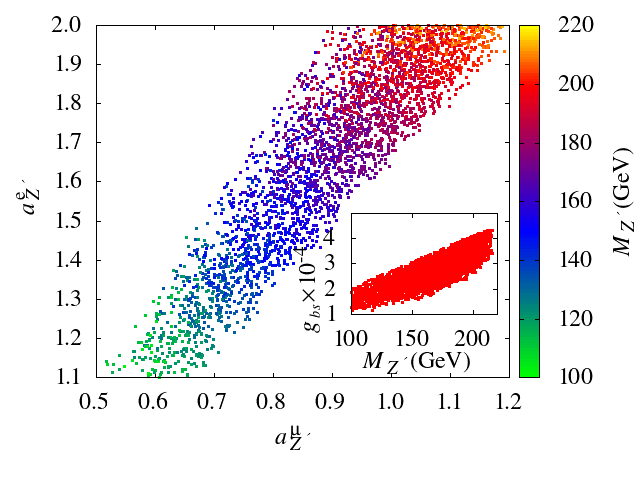}}
\caption{The plots are generated from the results of the package {\tt flavio}. Similar plots have been presented in Fig.~\ref{fig:all}, Fig.~\ref{fig:CCFR} and Fig.~\ref{fig:aZeaZmuMZpgbs_plus_g2} respectively, but obtained from the results of our in-house code.} 
\label{fig:flavio_all}
\end{center}
\end{figure}

Using the package {\tt flavio} the relevant plots have been generated and presented in Figs.~\ref{ab_flavio}, \ref{MZpgbs_flavio}, \ref{fig:flavio_CCFR} and \ref{fig:flavio_pos}. These may be compared with Figs.~\ref{ab}, \ref{MZpgbs}, \ref{fig:CCFR} and \ref{fig:aZeaZmuMZpgbs_plus_g2} respectively. Admittedly there is numerical differences which remains below $\sim 7\%$ in the region of interest. However, the qualitative nature of the results obtained remain consistent with each other.






}

\end{appendix}

\newpage

\providecommand{\href}[2]{#2}\begingroup\raggedright\endgroup

\begin{thebibliography}{100}

\bibitem{Muong-2:2021ojo}
{\scshape Muon g-2} collaboration, B.~Abi et~al., \emph{{Measurement of the
  Positive Muon Anomalous Magnetic Moment to 0.46 ppm}},
  \href{https://doi.org/10.1103/PhysRevLett.126.141801}{\emph{Phys. Rev. Lett.}
  {\bfseries 126} (2021) 141801},
  [\href{https://arxiv.org/abs/2104.03281}{{\ttfamily 2104.03281}}].

\bibitem{Borsanyi:2020mff}
S.~Borsanyi et~al., \emph{{Leading hadronic contribution to the muon magnetic
  moment from lattice QCD}},
  \href{https://doi.org/10.1038/s41586-021-03418-1}{\emph{Nature} {\bfseries
  593} (2021) 51--55}, [\href{https://arxiv.org/abs/2002.12347}{{\ttfamily
  2002.12347}}].

\bibitem{Parker:2018vye}
R.~H. Parker, C.~Yu, W.~Zhong, B.~Estey and H.~M\"uller, \emph{{Measurement of
  the fine-structure constant as a test of the Standard Model}},
  \href{https://doi.org/10.1126/science.aap7706}{\emph{Science} {\bfseries 360}
  (2018) 191}, [\href{https://arxiv.org/abs/1812.04130}{{\ttfamily
  1812.04130}}].

\bibitem{Hanneke:2008tm}
D.~Hanneke, S.~Fogwell and G.~Gabrielse, \emph{{New Measurement of the Electron
  Magnetic Moment and the Fine Structure Constant}},
  \href{https://doi.org/10.1103/PhysRevLett.100.120801}{\emph{Phys. Rev. Lett.}
  {\bfseries 100} (2008) 120801},
  [\href{https://arxiv.org/abs/0801.1134}{{\ttfamily 0801.1134}}].

\bibitem{SHIFMAN1979385}
M.~Shifman, A.~Vainshtein and V.~Zakharov, \emph{Qcd and resonance physics.
  theoretical foundations},
  \href{https://doi.org/https://doi.org/10.1016/0550-3213(79)90022-1}{\emph{Nuclear
  Physics B} {\bfseries 147} (1979) 385--447}.

\bibitem{Colangelo:2000dp}
P.~Colangelo and A.~Khodjamirian, \emph{{QCD sum rules, a modern perspective}},
   \href{https://arxiv.org/abs/hep-ph/0010175}{{\ttfamily hep-ph/0010175}}.

\bibitem{Descotes-Genon:2015uva}
S.~Descotes-Genon, L.~Hofer, J.~Matias and J.~Virto, \emph{{Global analysis of
  $b\to s\ell\ell$ anomalies}},
  \href{https://doi.org/10.1007/JHEP06(2016)092}{\emph{JHEP} {\bfseries 06}
  (2016) 092}, [\href{https://arxiv.org/abs/1510.04239}{{\ttfamily
  1510.04239}}].

\bibitem{Bordone:2016gaq}
M.~Bordone, G.~Isidori and A.~Pattori, \emph{{On the Standard Model predictions
  for $R_K$ and $R_{K^*}$}},
  \href{https://doi.org/10.1140/epjc/s10052-016-4274-7}{\emph{Eur. Phys. J. C}
  {\bfseries 76} (2016) 440},
  [\href{https://arxiv.org/abs/1605.07633}{{\ttfamily 1605.07633}}].

\bibitem{Capdevila:2017bsm}
B.~Capdevila, A.~Crivellin, S.~Descotes-Genon, J.~Matias and J.~Virto,
  \emph{{Patterns of New Physics in $b\to s\ell^+\ell^-$ transitions in the
  light of recent data}},
  \href{https://doi.org/10.1007/JHEP01(2018)093}{\emph{JHEP} {\bfseries 01}
  (2018) 093}, [\href{https://arxiv.org/abs/1704.05340}{{\ttfamily
  1704.05340}}].

\bibitem{LHCb:2021trn}
{\scshape LHCb} collaboration, R.~Aaij et~al., \emph{{Test of lepton
  universality in beauty-quark decays}},
  \href{https://doi.org/10.1038/s41567-021-01478-8}{\emph{Nature Phys.}
  {\bfseries 18} (2022) 277--282},
  [\href{https://arxiv.org/abs/2103.11769}{{\ttfamily 2103.11769}}].

\bibitem{Aaij:2017vbb}
{\scshape LHCb} collaboration, R.~Aaij et~al., \emph{{Test of lepton
  universality with $B^{0} \rightarrow K^{*0}\ell^{+}\ell^{-}$ decays}},
  \href{https://doi.org/10.1007/JHEP08(2017)055}{\emph{JHEP} {\bfseries 08}
  (2017) 055}, [\href{https://arxiv.org/abs/1705.05802}{{\ttfamily
  1705.05802}}].

\bibitem{Gauld:2013qba}
R.~Gauld, F.~Goertz and U.~Haisch, \emph{{On minimal $Z'$ explanations of the
  $B\to K^*\mu^+\mu^-$ anomaly}},
  \href{https://doi.org/10.1103/PhysRevD.89.015005}{\emph{Phys. Rev. D}
  {\bfseries 89} (2014) 015005},
  [\href{https://arxiv.org/abs/1308.1959}{{\ttfamily 1308.1959}}].

\bibitem{Glashow:2014iga}
S.~L. Glashow, D.~Guadagnoli and K.~Lane, \emph{{Lepton Flavor Violation in $B$
  Decays?}}, \href{https://doi.org/10.1103/PhysRevLett.114.091801}{\emph{Phys.
  Rev. Lett.} {\bfseries 114} (2015) 091801},
  [\href{https://arxiv.org/abs/1411.0565}{{\ttfamily 1411.0565}}].

\bibitem{Bhattacharya:2014wla}
B.~Bhattacharya, A.~Datta, D.~London and S.~Shivashankara, \emph{{Simultaneous
  Explanation of the $R_K$ and $R(D^{(*)})$ Puzzles}},
  \href{https://doi.org/10.1016/j.physletb.2015.02.011}{\emph{Phys. Lett. B}
  {\bfseries 742} (2015) 370--374},
  [\href{https://arxiv.org/abs/1412.7164}{{\ttfamily 1412.7164}}].

\bibitem{Crivellin:2015mga}
A.~Crivellin, G.~D'Ambrosio and J.~Heeck, \emph{{Explaining
  $h\to\mu^\pm\tau^\mp$, $B\to K^* \mu^+\mu^-$ and $B\to K \mu^+\mu^-/B\to K
  e^+e^-$ in a two-Higgs-doublet model with gauged $L_\mu-L_\tau$}},
  \href{https://doi.org/10.1103/PhysRevLett.114.151801}{\emph{Phys. Rev. Lett.}
  {\bfseries 114} (2015) 151801},
  [\href{https://arxiv.org/abs/1501.00993}{{\ttfamily 1501.00993}}].

\bibitem{Crivellin:2015era}
A.~Crivellin, L.~Hofer, J.~Matias, U.~Nierste, S.~Pokorski and J.~Rosiek,
  \emph{{Lepton-flavour violating $B$ decays in generic $Z'$ models}},
  \href{https://doi.org/10.1103/PhysRevD.92.054013}{\emph{Phys. Rev. D}
  {\bfseries 92} (2015) 054013},
  [\href{https://arxiv.org/abs/1504.07928}{{\ttfamily 1504.07928}}].

\bibitem{Celis:2015ara}
A.~Celis, J.~Fuentes-Martin, M.~Jung and H.~Serodio, \emph{{Family nonuniversal
  Z' models with protected flavor-changing interactions}},
  \href{https://doi.org/10.1103/PhysRevD.92.015007}{\emph{Phys. Rev. D}
  {\bfseries 92} (2015) 015007},
  [\href{https://arxiv.org/abs/1505.03079}{{\ttfamily 1505.03079}}].

\bibitem{Sierra:2015fma}
D.~Aristizabal~Sierra, F.~Staub and A.~Vicente, \emph{{Shedding light on the
  $b\to s$ anomalies with a dark sector}},
  \href{https://doi.org/10.1103/PhysRevD.92.015001}{\emph{Phys. Rev. D}
  {\bfseries 92} (2015) 015001},
  [\href{https://arxiv.org/abs/1503.06077}{{\ttfamily 1503.06077}}].

\bibitem{Belanger:2015nma}
G.~B\'elanger, C.~Delaunay and S.~Westhoff, \emph{{A Dark Matter Relic From
  Muon Anomalies}},
  \href{https://doi.org/10.1103/PhysRevD.92.055021}{\emph{Phys. Rev. D}
  {\bfseries 92} (2015) 055021},
  [\href{https://arxiv.org/abs/1507.06660}{{\ttfamily 1507.06660}}].

\bibitem{Gripaios:2015gra}
B.~Gripaios, M.~Nardecchia and S.~A. Renner, \emph{{Linear flavour violation
  and anomalies in B physics}},
  \href{https://doi.org/10.1007/JHEP06(2016)083}{\emph{JHEP} {\bfseries 06}
  (2016) 083}, [\href{https://arxiv.org/abs/1509.05020}{{\ttfamily
  1509.05020}}].

\bibitem{Allanach:2015gkd}
B.~Allanach, F.~S. Queiroz, A.~Strumia and S.~Sun, \emph{{$Z^\prime$ models for the
  LHCb and $g-2$ muon anomalies}},
  \href{https://doi.org/10.1103/PhysRevD.93.055045}{\emph{Phys. Rev. D}
  {\bfseries 93} (2016) 055045},
  [\href{https://arxiv.org/abs/1511.07447}{{\ttfamily 1511.07447}}].

\bibitem{Fuyuto:2015gmk}
K.~Fuyuto, W.-S. Hou and M.~Kohda, \emph{{Z' -induced FCNC decays of top,
  beauty, and strange quarks}},
  \href{https://doi.org/10.1103/PhysRevD.93.054021}{\emph{Phys. Rev. D}
  {\bfseries 93} (2016) 054021},
  [\href{https://arxiv.org/abs/1512.09026}{{\ttfamily 1512.09026}}].

\bibitem{Chiang:2016qov}
C.-W. Chiang, X.-G. He and G.~Valencia, \emph{{$Z^\prime$ model for
  b\textrightarrow{}s\ensuremath{\ell}$\overline{\ell}$ flavor anomalies}},
  \href{https://doi.org/10.1103/PhysRevD.93.074003}{\emph{Phys. Rev. D}
  {\bfseries 93} (2016) 074003},
  [\href{https://arxiv.org/abs/1601.07328}{{\ttfamily 1601.07328}}].

\bibitem{Boucenna:2016wpr}
S.~M. Boucenna, A.~Celis, J.~Fuentes-Martin, A.~Vicente and J.~Virto,
  \emph{{Non-abelian gauge extensions for B-decay anomalies}},
  \href{https://doi.org/10.1016/j.physletb.2016.06.067}{\emph{Phys. Lett. B}
  {\bfseries 760} (2016) 214--219},
  [\href{https://arxiv.org/abs/1604.03088}{{\ttfamily 1604.03088}}].

\bibitem{Boucenna:2016qad}
S.~M. Boucenna, A.~Celis, J.~Fuentes-Martin, A.~Vicente and J.~Virto,
  \emph{{Phenomenology of an $SU(2) \times SU(2) \times U(1)$ model with
  lepton-flavour non-universality}},
  \href{https://doi.org/10.1007/JHEP12(2016)059}{\emph{JHEP} {\bfseries 12}
  (2016) 059}, [\href{https://arxiv.org/abs/1608.01349}{{\ttfamily
  1608.01349}}].

\bibitem{Celis:2016ayl}
A.~Celis, W.-Z. Feng and M.~Vollmann, \emph{{Dirac dark matter and $b \to s
  \ell^+ \ell^-$ with $\mathrm{U(1)}$ gauge symmetry}},
  \href{https://doi.org/10.1103/PhysRevD.95.035018}{\emph{Phys. Rev. D}
  {\bfseries 95} (2017) 035018},
  [\href{https://arxiv.org/abs/1608.03894}{{\ttfamily 1608.03894}}].

\bibitem{Altmannshofer:2016jzy}
W.~Altmannshofer, S.~Gori, S.~Profumo and F.~S. Queiroz, \emph{{Explaining dark
  matter and B decay anomalies with an $L_\mu - L_\tau$ model}},
  \href{https://doi.org/10.1007/JHEP12(2016)106}{\emph{JHEP} {\bfseries 12}
  (2016) 106}, [\href{https://arxiv.org/abs/1609.04026}{{\ttfamily
  1609.04026}}].

\bibitem{Bhattacharya:2016mcc}
B.~Bhattacharya, A.~Datta, J.-P. Gu\'evin, D.~London and R.~Watanabe,
  \emph{{Simultaneous Explanation of the $R_K$ and $R_{D^{(*)}}$ Puzzles: a
  Model Analysis}}, \href{https://doi.org/10.1007/JHEP01(2017)015}{\emph{JHEP}
  {\bfseries 01} (2017) 015},
  [\href{https://arxiv.org/abs/1609.09078}{{\ttfamily 1609.09078}}].

\bibitem{Crivellin:2016ejn}
A.~Crivellin, J.~Fuentes-Martin, A.~Greljo and G.~Isidori, \emph{{Lepton Flavor
  Non-Universality in B decays from Dynamical Yukawas}},
  \href{https://doi.org/10.1016/j.physletb.2016.12.057}{\emph{Phys. Lett. B}
  {\bfseries 766} (2017) 77--85},
  [\href{https://arxiv.org/abs/1611.02703}{{\ttfamily 1611.02703}}].

\bibitem{Becirevic:2016zri}
D.~Be\v{c}irevi\'c, O.~Sumensari and R.~Zukanovich~Funchal, \emph{{Lepton
  flavor violation in exclusive $b\rightarrow s$ decays}},
  \href{https://doi.org/10.1140/epjc/s10052-016-3985-0}{\emph{Eur. Phys. J. C}
  {\bfseries 76} (2016) 134},
  [\href{https://arxiv.org/abs/1602.00881}{{\ttfamily 1602.00881}}].

\bibitem{GarciaGarcia:2016nvr}
I.~Garcia~Garcia, \emph{{LHCb anomalies from a natural perspective}},
  \href{https://doi.org/10.1007/JHEP03(2017)040}{\emph{JHEP} {\bfseries 03}
  (2017) 040}, [\href{https://arxiv.org/abs/1611.03507}{{\ttfamily
  1611.03507}}].

\bibitem{Bhatia:2017tgo}
D.~Bhatia, S.~Chakraborty and A.~Dighe, \emph{{Neutrino mixing and $R_K$
  anomaly in U(1)$_X$ models: a bottom-up approach}},
  \href{https://doi.org/10.1007/JHEP03(2017)117}{\emph{JHEP} {\bfseries 03}
  (2017) 117}, [\href{https://arxiv.org/abs/1701.05825}{{\ttfamily
  1701.05825}}].

\bibitem{Ko:2017yrd}
P.~Ko, T.~Nomura and H.~Okada, \emph{{Explaining $B\to K^{(*)}\ell^+ \ell^-$
  anomaly by radiatively induced coupling in $U(1)_{\mu-\tau}$ gauge
  symmetry}}, \href{https://doi.org/10.1103/PhysRevD.95.111701}{\emph{Phys.
  Rev. D} {\bfseries 95} (2017) 111701},
  [\href{https://arxiv.org/abs/1702.02699}{{\ttfamily 1702.02699}}].

\bibitem{Chen:2017usq}
C.-H. Chen and T.~Nomura, \emph{{Penguin $b \to s\ell'^+ \ell'^-$ and $B$-meson
  anomalies in a gauged ${L_\mu -L_\tau}$}},
  \href{https://doi.org/10.1016/j.physletb.2017.12.062}{\emph{Phys. Lett. B}
  {\bfseries 777} (2018) 420--427},
  [\href{https://arxiv.org/abs/1707.03249}{{\ttfamily 1707.03249}}].

\bibitem{Baek:2017sew}
S.~Baek, \emph{{Dark matter contribution to $b\to s \mu^+ \mu^-$ anomaly in
  local $U(1)_{L_\mu-L_\tau}$ model}},
  \href{https://doi.org/10.1016/j.physletb.2018.04.012}{\emph{Phys. Lett. B}
  {\bfseries 781} (2018) 376--382},
  [\href{https://arxiv.org/abs/1707.04573}{{\ttfamily 1707.04573}}].

\bibitem{King:2017anf}
S.~F. King, \emph{{Flavourful $Z^\prime$ models for $ {R}_{K^{\left(\ast
  \right)}} $}}, \href{https://doi.org/10.1007/JHEP08(2017)019}{\emph{JHEP}
  {\bfseries 08} (2017) 019},
  [\href{https://arxiv.org/abs/1706.06100}{{\ttfamily 1706.06100}}].

\bibitem{King:2018fcg}
S.~F. King, \emph{{$ {R}_{K^{\left(*\right)}} $ and the origin of Yukawa
  couplings}}, \href{https://doi.org/10.1007/JHEP09(2018)069}{\emph{JHEP}
  {\bfseries 09} (2018) 069},
  [\href{https://arxiv.org/abs/1806.06780}{{\ttfamily 1806.06780}}].

\bibitem{Dasgupta:2018nzt}
S.~Dasgupta, U.~K. Dey, T.~Jha and T.~S. Ray, \emph{{Status of a flavor-maximal
  nonminimal universal extra dimension model}},
  \href{https://doi.org/10.1103/PhysRevD.98.055006}{\emph{Phys. Rev. D}
  {\bfseries 98} (2018) 055006},
  [\href{https://arxiv.org/abs/1801.09722}{{\ttfamily 1801.09722}}].

\bibitem{Biswas:2019twf}
A.~Biswas and A.~Shaw, \emph{{Reconciling dark matter, $R_{K^{(*)}}$ anomalies
  and $(g-2)_{\mu}$ in an ${L_{\mu}-L_{\tau}}$ scenario}},
  \href{https://doi.org/10.1007/JHEP05(2019)165}{\emph{JHEP} {\bfseries 05}
  (2019) 165}, [\href{https://arxiv.org/abs/1903.08745}{{\ttfamily
  1903.08745}}].

\bibitem{Dwivedi:2019uqd}
S.~Dwivedi, D.~Kumar~Ghosh, A.~Falkowski and N.~Ghosh, \emph{{Associated
  $Z^\prime$ production in the flavorful $U(1)$ scenario for $R_{K^{(*)}}$}},
  \href{https://doi.org/10.1140/epjc/s10052-020-7810-4}{\emph{Eur. Phys. J. C}
  {\bfseries 80} (2020) 263},
  [\href{https://arxiv.org/abs/1908.03031}{{\ttfamily 1908.03031}}].

\bibitem{CarcamoHernandez:2019ydc}
A.~E. C\'arcamo~Hern\'andez, S.~F. King, H.~Lee and S.~J. Rowley, \emph{{Is it
  possible to explain the muon and electron $g-2$ in a $Z^\prime$ model?}},
  \href{https://doi.org/10.1103/PhysRevD.101.115016}{\emph{Phys. Rev. D}
  {\bfseries 101} (2020) 115016},
  [\href{https://arxiv.org/abs/1910.10734}{{\ttfamily 1910.10734}}].

\bibitem{Bodas:2021fsy}
A.~Bodas, R.~Coy and S.~J.~D. King, \emph{{Solving the electron and muon $g-2$
  anomalies in $Z'$ models}},
  \href{https://doi.org/10.1140/epjc/s10052-021-09850-x}{\emph{Eur. Phys. J. C}
  {\bfseries 81} (2021) 1065},
  [\href{https://arxiv.org/abs/2102.07781}{{\ttfamily 2102.07781}}].

\bibitem{Biswas:2021dan}
A.~Biswas and S.~Khan, \emph{{(g \ensuremath{-} 2)$_{e, \mu}$ and strongly
  interacting dark matter with collider implications}},
  \href{https://doi.org/10.1007/JHEP07(2022)037}{\emph{JHEP} {\bfseries 07}
  (2022) 037}, [\href{https://arxiv.org/abs/2112.08393}{{\ttfamily
  2112.08393}}].

\bibitem{Hiller:2014yaa}
G.~Hiller and M.~Schmaltz, \emph{{$R_K$ and future $b \to s \ell \ell$ physics
  beyond the standard model opportunities}},
  \href{https://doi.org/10.1103/PhysRevD.90.054014}{\emph{Phys. Rev. D}
  {\bfseries 90} (2014) 054014},
  [\href{https://arxiv.org/abs/1408.1627}{{\ttfamily 1408.1627}}].

\bibitem{Biswas:2014gga}
S.~Biswas, D.~Chowdhury, S.~Han and S.~J. Lee, \emph{{Explaining the lepton
  non-universality at the LHCb and CMS within a unified framework}},
  \href{https://doi.org/10.1007/JHEP02(2015)142}{\emph{JHEP} {\bfseries 02}
  (2015) 142}, [\href{https://arxiv.org/abs/1409.0882}{{\ttfamily 1409.0882}}].

\bibitem{Gripaios:2014tna}
B.~Gripaios, M.~Nardecchia and S.~A. Renner, \emph{{Composite leptoquarks and
  anomalies in $B$-meson decays}},
  \href{https://doi.org/10.1007/JHEP05(2015)006}{\emph{JHEP} {\bfseries 05}
  (2015) 006}, [\href{https://arxiv.org/abs/1412.1791}{{\ttfamily 1412.1791}}].

\bibitem{Sahoo:2015wya}
S.~Sahoo and R.~Mohanta, \emph{{Scalar leptoquarks and the rare $B$ meson
  decays}}, \href{https://doi.org/10.1103/PhysRevD.91.094019}{\emph{Phys. Rev.
  D} {\bfseries 91} (2015) 094019},
  [\href{https://arxiv.org/abs/1501.05193}{{\ttfamily 1501.05193}}].

\bibitem{Becirevic:2015asa}
D.~Be\v{c}irevi\'c, S.~Fajfer and N.~Ko\v{s}nik, \emph{{Lepton flavor
  nonuniversality in
  b\textrightarrow{}s\ensuremath{\ell}$^+$\ensuremath{\ell}$^-$ processes}},
  \href{https://doi.org/10.1103/PhysRevD.92.014016}{\emph{Phys. Rev. D}
  {\bfseries 92} (2015) 014016},
  [\href{https://arxiv.org/abs/1503.09024}{{\ttfamily 1503.09024}}].

\bibitem{Alonso:2015sja}
R.~Alonso, B.~Grinstein and J.~Martin~Camalich, \emph{{Lepton universality
  violation and lepton flavor conservation in $B$-meson decays}},
  \href{https://doi.org/10.1007/JHEP10(2015)184}{\emph{JHEP} {\bfseries 10}
  (2015) 184}, [\href{https://arxiv.org/abs/1505.05164}{{\ttfamily
  1505.05164}}].

\bibitem{Calibbi:2015kma}
L.~Calibbi, A.~Crivellin and T.~Ota, \emph{{Effective Field Theory Approach to
  $b\to s\ell\ell^{(')}$, $B\to K^{(*)}\nu\overline{\nu}$ and $B\to
  D^{(*)}\tau\nu$ with Third Generation Couplings}},
  \href{https://doi.org/10.1103/PhysRevLett.115.181801}{\emph{Phys. Rev. Lett.}
  {\bfseries 115} (2015) 181801},
  [\href{https://arxiv.org/abs/1506.02661}{{\ttfamily 1506.02661}}].

\bibitem{Huang:2015vpt}
W.~Huang and Y.-L. Tang, \emph{{Flavor anomalies at the LHC and the R-parity
  violating supersymmetric model extended with vectorlike particles}},
  \href{https://doi.org/10.1103/PhysRevD.92.094015}{\emph{Phys. Rev. D}
  {\bfseries 92} (2015) 094015},
  [\href{https://arxiv.org/abs/1509.08599}{{\ttfamily 1509.08599}}].

\bibitem{Pas:2015hca}
H.~P\"as and E.~Schumacher, \emph{{Common origin of $R_K$ and neutrino
  masses}}, \href{https://doi.org/10.1103/PhysRevD.92.114025}{\emph{Phys. Rev.
  D} {\bfseries 92} (2015) 114025},
  [\href{https://arxiv.org/abs/1510.08757}{{\ttfamily 1510.08757}}].

\bibitem{Bauer:2015knc}
M.~Bauer and M.~Neubert, \emph{{Minimal Leptoquark Explanation for the
  $R_{D^{(*)}}$ , $R_K$ , and $(g-2)_\mu$ Anomalies}},
  \href{https://doi.org/10.1103/PhysRevLett.116.141802}{\emph{Phys. Rev. Lett.}
  {\bfseries 116} (2016) 141802},
  [\href{https://arxiv.org/abs/1511.01900}{{\ttfamily 1511.01900}}].

\bibitem{Fajfer:2015ycq}
S.~Fajfer and N.~Ko\v{s}nik, \emph{{Vector leptoquark resolution of $R_K$ and
  $R_{D^{(*)}}$ puzzles}},
  \href{https://doi.org/10.1016/j.physletb.2016.02.018}{\emph{Phys. Lett. B}
  {\bfseries 755} (2016) 270--274},
  [\href{https://arxiv.org/abs/1511.06024}{{\ttfamily 1511.06024}}].

\bibitem{Barbieri:2015yvd}
R.~Barbieri, G.~Isidori, A.~Pattori and F.~Senia, \emph{{Anomalies in
  $B$-decays and $U(2)$ flavour symmetry}},
  \href{https://doi.org/10.1140/epjc/s10052-016-3905-3}{\emph{Eur. Phys. J. C}
  {\bfseries 76} (2016) 67},
  [\href{https://arxiv.org/abs/1512.01560}{{\ttfamily 1512.01560}}].

\bibitem{Sahoo:2015pzk}
S.~Sahoo and R.~Mohanta, \emph{{Lepton flavor violating B meson decays via a
  scalar leptoquark}},
  \href{https://doi.org/10.1103/PhysRevD.93.114001}{\emph{Phys. Rev. D}
  {\bfseries 93} (2016) 114001},
  [\href{https://arxiv.org/abs/1512.04657}{{\ttfamily 1512.04657}}].

\bibitem{Dorsner:2016wpm}
I.~Dor\v{s}ner, S.~Fajfer, A.~Greljo, J.~F. Kamenik and N.~Ko\v{s}nik,
  \emph{{Physics of leptoquarks in precision experiments and at particle
  colliders}}, \href{https://doi.org/10.1016/j.physrep.2016.06.001}{\emph{Phys.
  Rept.} {\bfseries 641} (2016) 1--68},
  [\href{https://arxiv.org/abs/1603.04993}{{\ttfamily 1603.04993}}].

\bibitem{Sahoo:2016nvx}
S.~Sahoo and R.~Mohanta, \emph{{Effects of scalar leptoquark on semileptonic
  $\Lambda_b$ decays}},
  \href{https://doi.org/10.1088/1367-2630/18/9/093051}{\emph{New J. Phys.}
  {\bfseries 18} (2016) 093051},
  [\href{https://arxiv.org/abs/1607.04449}{{\ttfamily 1607.04449}}].

\bibitem{Das:2016vkr}
D.~Das, C.~Hati, G.~Kumar and N.~Mahajan, \emph{{Towards a unified explanation
  of $R_{D^{(\ast)}}$, $R_{K}$ and $(g-2)_{\mu}$ anomalies in a left-right
  model with leptoquarks}},
  \href{https://doi.org/10.1103/PhysRevD.94.055034}{\emph{Phys. Rev. D}
  {\bfseries 94} (2016) 055034},
  [\href{https://arxiv.org/abs/1605.06313}{{\ttfamily 1605.06313}}].

\bibitem{Chen:2016dip}
C.-H. Chen, T.~Nomura and H.~Okada, \emph{{Explanation of $B \to K^{(*)} \ell^+
  \ell^-$ and muon $g-2$, and implications at the LHC}},
  \href{https://doi.org/10.1103/PhysRevD.94.115005}{\emph{Phys. Rev. D}
  {\bfseries 94} (2016) 115005},
  [\href{https://arxiv.org/abs/1607.04857}{{\ttfamily 1607.04857}}].

\bibitem{Becirevic:2016oho}
D.~Be\v{c}irevi\'c, N.~Ko\v{s}nik, O.~Sumensari and R.~Zukanovich~Funchal,
  \emph{{Palatable Leptoquark Scenarios for Lepton Flavor Violation in
  Exclusive $b\to s\ell_1\ell_2$ modes}},
  \href{https://doi.org/10.1007/JHEP11(2016)035}{\emph{JHEP} {\bfseries 11}
  (2016) 035}, [\href{https://arxiv.org/abs/1608.07583}{{\ttfamily
  1608.07583}}].

\bibitem{Becirevic:2016yqi}
D.~Be\v{c}irevi\'c, S.~Fajfer, N.~Ko\v{s}nik and O.~Sumensari,
  \emph{{Leptoquark model to explain the $B$-physics anomalies, $R_K$ and
  $R_D$}}, \href{https://doi.org/10.1103/PhysRevD.94.115021}{\emph{Phys. Rev.
  D} {\bfseries 94} (2016) 115021},
  [\href{https://arxiv.org/abs/1608.08501}{{\ttfamily 1608.08501}}].

\bibitem{Sahoo:2016pet}
S.~Sahoo, R.~Mohanta and A.~K. Giri, \emph{{Explaining the $R_{K}$ and
  $R_{D^{(*)}}$ anomalies with vector leptoquarks}},
  \href{https://doi.org/10.1103/PhysRevD.95.035027}{\emph{Phys. Rev. D}
  {\bfseries 95} (2017) 035027},
  [\href{https://arxiv.org/abs/1609.04367}{{\ttfamily 1609.04367}}].

\bibitem{Barbieri:2016las}
R.~Barbieri, C.~W. Murphy and F.~Senia, \emph{{B-decay Anomalies in a Composite
  Leptoquark Model}},
  \href{https://doi.org/10.1140/epjc/s10052-016-4578-7}{\emph{Eur. Phys. J. C}
  {\bfseries 77} (2017) 8}, [\href{https://arxiv.org/abs/1611.04930}{{\ttfamily
  1611.04930}}].

\bibitem{Cox:2016epl}
P.~Cox, A.~Kusenko, O.~Sumensari and T.~T. Yanagida, \emph{{SU(5) Unification
  with TeV-scale Leptoquarks}},
  \href{https://doi.org/10.1007/JHEP03(2017)035}{\emph{JHEP} {\bfseries 03}
  (2017) 035}, [\href{https://arxiv.org/abs/1612.03923}{{\ttfamily
  1612.03923}}].

\bibitem{Ma:2001md}
E.~Ma, D.~P. Roy and S.~Roy, \emph{{Gauged L(mu) - L(tau) with large muon
  anomalous magnetic moment and the bimaximal mixing of neutrinos}},
  \href{https://doi.org/10.1016/S0370-2693(01)01428-9}{\emph{Phys. Lett. B}
  {\bfseries 525} (2002) 101--106},
  [\href{https://arxiv.org/abs/hep-ph/0110146}{{\ttfamily hep-ph/0110146}}].

\bibitem{Baek:2001kca}
S.~Baek, N.~G. Deshpande, X.~G. He and P.~Ko, \emph{{Muon anomalous g-2 and
  gauged L(muon) - L(tau) models}},
  \href{https://doi.org/10.1103/PhysRevD.64.055006}{\emph{Phys. Rev. D}
  {\bfseries 64} (2001) 055006},
  [\href{https://arxiv.org/abs/hep-ph/0104141}{{\ttfamily hep-ph/0104141}}].

\bibitem{Heeck:2011wj}
J.~Heeck and W.~Rodejohann, \emph{{Gauged $L_\mu - L_\tau$ Symmetry at the
  Electroweak Scale}},
  \href{https://doi.org/10.1103/PhysRevD.84.075007}{\emph{Phys. Rev. D}
  {\bfseries 84} (2011) 075007},
  [\href{https://arxiv.org/abs/1107.5238}{{\ttfamily 1107.5238}}].

\bibitem{Harigaya:2013twa}
K.~Harigaya, T.~Igari, M.~M. Nojiri, M.~Takeuchi and K.~Tobe, \emph{{Muon g-2
  and LHC phenomenology in the $L_\mu-L_\tau$ gauge symmetric model}},
  \href{https://doi.org/10.1007/JHEP03(2014)105}{\emph{JHEP} {\bfseries 03}
  (2014) 105}, [\href{https://arxiv.org/abs/1311.0870}{{\ttfamily 1311.0870}}].

\bibitem{Altmannshofer:2016brv}
W.~Altmannshofer, C.-Y. Chen, P.~S. Bhupal~Dev and A.~Soni, \emph{{Lepton
  flavor violating Z' explanation of the muon anomalous magnetic moment}},
  \href{https://doi.org/10.1016/j.physletb.2016.09.046}{\emph{Phys. Lett. B}
  {\bfseries 762} (2016) 389--398},
  [\href{https://arxiv.org/abs/1607.06832}{{\ttfamily 1607.06832}}].

\bibitem{Biswas:2016yan}
A.~Biswas, S.~Choubey and S.~Khan, \emph{{Neutrino Mass, Dark Matter and
  Anomalous Magnetic Moment of Muon in a $U(1)_{L_{\mu}-L_{\tau}}$ Model}},
  \href{https://doi.org/10.1007/JHEP09(2016)147}{\emph{JHEP} {\bfseries 09}
  (2016) 147}, [\href{https://arxiv.org/abs/1608.04194}{{\ttfamily
  1608.04194}}].

\bibitem{Biswas:2016yjr}
A.~Biswas, S.~Choubey and S.~Khan, \emph{{FIMP and Muon ($g-2$) in a
  U$(1)_{L_{\mu}-L_{\tau}}$ Model}},
  \href{https://doi.org/10.1007/JHEP02(2017)123}{\emph{JHEP} {\bfseries 02}
  (2017) 123}, [\href{https://arxiv.org/abs/1612.03067}{{\ttfamily
  1612.03067}}].

\bibitem{Banerjee:2018eaf}
H.~Banerjee, P.~Byakti and S.~Roy, \emph{{Supersymmetric gauged
  U(1)$_{L_{\mu}-L_{\tau}}$ model for neutrinos and the muon $(g-2)$ anomaly}},
  \href{https://doi.org/10.1103/PhysRevD.98.075022}{\emph{Phys. Rev. D}
  {\bfseries 98} (2018) 075022},
  [\href{https://arxiv.org/abs/1805.04415}{{\ttfamily 1805.04415}}].

\bibitem{Huang:2020ris}
D.~Huang, A.~P. Morais and R.~Santos, \emph{{Anomalies in $B$-meson decays and
  the muon $g-2$ from dark loops}},
  \href{https://doi.org/10.1103/PhysRevD.102.075009}{\emph{Phys. Rev. D}
  {\bfseries 102} (2020) 075009},
  [\href{https://arxiv.org/abs/2007.05082}{{\ttfamily 2007.05082}}].

\bibitem{Dinh:2020inx}
S.~Q. Dinh and H.~M. Tran, \emph{{Muon g-2 and semileptonic B decays in the
  B\'elanger-Delaunay-Westhoff model with gauge kinetic mixing}},
  \href{https://doi.org/10.1103/PhysRevD.104.115009}{\emph{Phys. Rev. D}
  {\bfseries 104} (2021) 115009},
  [\href{https://arxiv.org/abs/2011.07182}{{\ttfamily 2011.07182}}].

\bibitem{Chakraborti:2021kkr}
M.~Chakraborti, S.~Heinemeyer and I.~Saha, \emph{{Improved ${(g-2)_\mu }$
  measurements and wino/higgsino dark matter}},
  \href{https://doi.org/10.1140/epjc/s10052-021-09814-1}{\emph{Eur. Phys. J. C}
  {\bfseries 81} (2021) 1069},
  [\href{https://arxiv.org/abs/2103.13403}{{\ttfamily 2103.13403}}].

\bibitem{Chakraborti:2021dli}
M.~Chakraborti, S.~Heinemeyer and I.~Saha, \emph{{The new
  \textquotedblleft{}MUON G-2\textquotedblright{} result and supersymmetry}},
  \href{https://doi.org/10.1140/epjc/s10052-021-09900-4}{\emph{Eur. Phys. J. C}
  {\bfseries 81} (2021) 1114},
  [\href{https://arxiv.org/abs/2104.03287}{{\ttfamily 2104.03287}}].

\bibitem{Arcadi:2021cwg}
G.~Arcadi, L.~Calibbi, M.~Fedele and F.~Mescia, \emph{{Muon $g-2$ and
  $B$-anomalies from Dark Matter}},
  \href{https://doi.org/10.1103/PhysRevLett.127.061802}{\emph{Phys. Rev. Lett.}
  {\bfseries 127} (2021) 061802},
  [\href{https://arxiv.org/abs/2104.03228}{{\ttfamily 2104.03228}}].

\bibitem{Alvarado:2021nxy}
J.~S. Alvarado, S.~F. Mantilla, R.~Martinez and F.~Ochoa, \emph{{A
  non-universal $U(1)_{X}$ extension to the Standard Model to study the $B$
  meson anomaly and muon $g-2$}},
  \href{https://arxiv.org/abs/2105.04715}{{\ttfamily 2105.04715}}.

\bibitem{Davighi:2021oel}
J.~Davighi, \emph{{Anomalous Z' bosons for anomalous B decays}},
  \href{https://doi.org/10.1007/JHEP08(2021)101}{\emph{JHEP} {\bfseries 08}
  (2021) 101}, [\href{https://arxiv.org/abs/2105.06918}{{\ttfamily
  2105.06918}}].

\bibitem{Darme:2021qzw}
L.~Darm\'e, M.~Fedele, K.~Kowalska and E.~M. Sessolo, \emph{{Flavour anomalies
  and the muon g \ensuremath{-} 2 from feebly interacting particles}},
  \href{https://doi.org/10.1007/JHEP03(2022)085}{\emph{JHEP} {\bfseries 03}
  (2022) 085}, [\href{https://arxiv.org/abs/2106.12582}{{\ttfamily
  2106.12582}}].

\bibitem{Lee:2021ndf}
J.-P. Lee, \emph{{$R(K^{(*)})$ with vector unparticles}},
  \href{https://arxiv.org/abs/2106.12795}{{\ttfamily 2106.12795}}.

\bibitem{Greljo:2021npi}
A.~Greljo, Y.~Soreq, P.~Stangl, A.~E. Thomsen and J.~Zupan, \emph{{Muonic force
  behind flavor anomalies}},
  \href{https://doi.org/10.1007/JHEP04(2022)151}{\emph{JHEP} {\bfseries 04}
  (2022) 151}, [\href{https://arxiv.org/abs/2107.07518}{{\ttfamily
  2107.07518}}].

\bibitem{Wang:2021uqz}
X.~Wang, \emph{{Muon (g \ensuremath{-} 2) and flavor puzzles in the
  U(1)$_{X}$-gauged leptoquark model}},
  \href{https://doi.org/10.1007/JHEP08(2022)243}{\emph{JHEP} {\bfseries 08}
  (2022) 243}, [\href{https://arxiv.org/abs/2108.01279}{{\ttfamily
  2108.01279}}].

\bibitem{Navarro:2021sfb}
M.~F. Navarro and S.~F. King, \emph{{Fermiophobic $Z^\prime$ model for simultaneously
  explaining the muon anomalies RK(*) and (g-2)\ensuremath{\mu}}},
  \href{https://doi.org/10.1103/PhysRevD.105.035015}{\emph{Phys. Rev. D}
  {\bfseries 105} (2022) 035015},
  [\href{https://arxiv.org/abs/2109.08729}{{\ttfamily 2109.08729}}].

\bibitem{Bause:2021prv}
R.~Bause, G.~Hiller, T.~H\"ohne, D.~F. Litim and T.~Steudtner,
  \emph{{B-anomalies from flavorful U(1)$'$ extensions, safely}},
  \href{https://doi.org/10.1140/epjc/s10052-021-09957-1}{\emph{Eur. Phys. J. C}
  {\bfseries 82} (2022) 42},
  [\href{https://arxiv.org/abs/2109.06201}{{\ttfamily 2109.06201}}].

\bibitem{Ko:2021lpx}
P.~Ko, T.~Nomura and H.~Okada, \emph{{Muon $g-2$, $B\to K^{(*)}\mu^+ \mu^-$
  anomalies, and leptophilic dark matter in $U(1)_{\mu-\tau}$ gauge symmetry}},
   \href{https://arxiv.org/abs/2110.10513}{{\ttfamily 2110.10513}}.


\bibitem{Zyla:2020zbs}
{\scshape Particle Data Group} collaboration, P.~A. Zyla et~al., \emph{{Review
  of Particle Physics}},
  \href{https://doi.org/10.1093/ptep/ptaa104}{\emph{PTEP} {\bfseries 2020}
  (2020) 083C01}.

\bibitem{LHCb:2020gog}
{\scshape LHCb} collaboration, R.~Aaij et~al., \emph{{Angular Analysis of the
  $B^{+}\rightarrow K^{\ast+}\mu^{+}\mu^{-}$ Decay}},
  \href{https://doi.org/10.1103/PhysRevLett.126.161802}{\emph{Phys. Rev. Lett.}
  {\bfseries 126} (2021) 161802},
  [\href{https://arxiv.org/abs/2012.13241}{{\ttfamily 2012.13241}}].

\bibitem{LHCb:2020pcv}
{\scshape LHCb} collaboration, R.~Aaij et~al., \emph{{Search for the Rare
  Decays $B^0_s\to e^+e^-$ and $B^0\to e^+e^-$}},
  \href{https://doi.org/10.1103/PhysRevLett.124.211802}{\emph{Phys. Rev. Lett.}
  {\bfseries 124} (2020) 211802},
  [\href{https://arxiv.org/abs/2003.03999}{{\ttfamily 2003.03999}}].

\bibitem{LHCb:2013pra}
{\scshape LHCb} collaboration, R.~Aaij et~al., \emph{{Measurement of the $B^0
  \rightarrow K^{*0}e^+e^-$ branching fraction at low dilepton mass}},
  \href{https://doi.org/10.1007/JHEP05(2013)159}{\emph{JHEP} {\bfseries 05}
  (2013) 159}, [\href{https://arxiv.org/abs/1304.3035}{{\ttfamily 1304.3035}}].

\bibitem{LHCb:2014vgu}
{\scshape LHCb} collaboration, R.~Aaij et~al., \emph{{Test of lepton
  universality using $B^{+}\rightarrow K^{+}\ell^{+}\ell^{-}$ decays}},
  \href{https://doi.org/10.1103/PhysRevLett.113.151601}{\emph{Phys. Rev. Lett.}
  {\bfseries 113} (2014) 151601},
  [\href{https://arxiv.org/abs/1406.6482}{{\ttfamily 1406.6482}}].

\bibitem{Belle:2017oht}
{\scshape Belle} collaboration, J.~Grygier et~al., \emph{{Search for
  $\boldsymbol{B\to h\nu\bar{\nu}}$ decays with semileptonic tagging at
  Belle}}, \href{https://doi.org/10.1103/PhysRevD.96.091101}{\emph{Phys. Rev.
  D} {\bfseries 96} (2017) 091101},
  [\href{https://arxiv.org/abs/1702.03224}{{\ttfamily 1702.03224}}].

\bibitem{Mishra:1991bv}
{\scshape CCFR} collaboration, S.~R. Mishra et~al., \emph{{Neutrino tridents
  and W Z interference}},
  \href{https://doi.org/10.1103/PhysRevLett.66.3117}{\emph{Phys. Rev. Lett.}
  {\bfseries 66} (1991) 3117--3120}.

\bibitem{Gninenko:2001hx}
S.~N. Gninenko and N.~V. Krasnikov, \emph{{The Muon anomalous magnetic moment
  and a new light gauge boson}},
  \href{https://doi.org/10.1016/S0370-2693(01)00693-1}{\emph{Phys. Lett. B}
  {\bfseries 513} (2001) 119},
  [\href{https://arxiv.org/abs/hep-ph/0102222}{{\ttfamily hep-ph/0102222}}].

\bibitem{Aoyama:2020ynm}
T.~Aoyama et~al., \emph{{The anomalous magnetic moment of the muon in the
  Standard Model}},
  \href{https://doi.org/10.1016/j.physrep.2020.07.006}{\emph{Phys. Rept.}
  {\bfseries 887} (2020) 1--166},
  [\href{https://arxiv.org/abs/2006.04822}{{\ttfamily 2006.04822}}].

\bibitem{Aoyama:2017uqe}
T.~Aoyama, T.~Kinoshita and M.~Nio, \emph{{Revised and Improved Value of the
  QED Tenth-Order Electron Anomalous Magnetic Moment}},
  \href{https://doi.org/10.1103/PhysRevD.97.036001}{\emph{Phys. Rev. D}
  {\bfseries 97} (2018) 036001},
  [\href{https://arxiv.org/abs/1712.06060}{{\ttfamily 1712.06060}}].

\bibitem{Hiller:2019mou}
G.~Hiller, C.~Hormigos-Feliu, D.~F. Litim and T.~Steudtner, \emph{{Anomalous
  magnetic moments from asymptotic safety}},
  \href{https://doi.org/10.1103/PhysRevD.102.071901}{\emph{Phys. Rev. D}
  {\bfseries 102} (2020) 071901},
  [\href{https://arxiv.org/abs/1910.14062}{{\ttfamily 1910.14062}}].

\bibitem{Hiller:2003js}
G.~Hiller and F.~Kruger, \emph{{More model-independent analysis of $b \to s$
  processes}}, \href{https://doi.org/10.1103/PhysRevD.69.074020}{\emph{Phys.
  Rev. D} {\bfseries 69} (2004) 074020},
  [\href{https://arxiv.org/abs/hep-ph/0310219}{{\ttfamily hep-ph/0310219}}].

\bibitem{Isidori:2020acz}
G.~Isidori, S.~Nabeebaccus and R.~Zwicky, \emph{{QED corrections in $
  \overline{B}\to \overline{K}{\mathrm{\ell}}^{+}{\mathrm{\ell}}^{-} $ at the
  double-differential level}},
  \href{https://doi.org/10.1007/JHEP12(2020)104}{\emph{JHEP} {\bfseries 12}
  (2020) 104}, [\href{https://arxiv.org/abs/2009.00929}{{\ttfamily
  2009.00929}}].

\bibitem{Buchalla:1995vs}
G.~Buchalla, A.~J. Buras and M.~E. Lautenbacher, \emph{{Weak decays beyond
  leading logarithms}},
  \href{https://doi.org/10.1103/RevModPhys.68.1125}{\emph{Rev. Mod. Phys.}
  {\bfseries 68} (1996) 1125--1144},
  [\href{https://arxiv.org/abs/hep-ph/9512380}{{\ttfamily hep-ph/9512380}}].

\bibitem{Buras:2012dp}
A.~J. Buras, F.~De~Fazio, J.~Girrbach and M.~V. Carlucci, \emph{{The Anatomy of
  Quark Flavour Observables in 331 Models in the Flavour Precision Era}},
  \href{https://doi.org/10.1007/JHEP02(2013)023}{\emph{JHEP} {\bfseries 02}
  (2013) 023}, [\href{https://arxiv.org/abs/1211.1237}{{\ttfamily 1211.1237}}].

\bibitem{Altmannshofer:2014rta}
W.~Altmannshofer and D.~M. Straub, \emph{{New physics in $b\rightarrow s$
  transitions after LHC run 1}},
  \href{https://doi.org/10.1140/epjc/s10052-015-3602-7}{\emph{Eur. Phys. J. C}
  {\bfseries 75} (2015) 382},
  [\href{https://arxiv.org/abs/1411.3161}{{\ttfamily 1411.3161}}].

\bibitem{Bobeth:2008ij}
C.~Bobeth, G.~Hiller and G.~Piranishvili, \emph{{CP Asymmetries in bar $B \to
  \bar{K}^* (\to \bar{K} \pi) \bar{\ell} \ell$ and Untagged $\bar{B}_s$, $B_s
  \to \phi (\to K^{+} K^-) \bar{\ell} \ell$ Decays at NLO}},
  \href{https://doi.org/10.1088/1126-6708/2008/07/106}{\emph{JHEP} {\bfseries
  07} (2008) 106}, [\href{https://arxiv.org/abs/0805.2525}{{\ttfamily
  0805.2525}}].

\bibitem{Matias:2012xw}
J.~Matias, F.~Mescia, M.~Ramon and J.~Virto, \emph{{Complete Anatomy of
  $\bar{B}_d -> \bar{K}^{* 0} (-> K \pi)l^+l^-$ and its angular distribution}},
  \href{https://doi.org/10.1007/JHEP04(2012)104}{\emph{JHEP} {\bfseries 04}
  (2012) 104}, [\href{https://arxiv.org/abs/1202.4266}{{\ttfamily 1202.4266}}].

\bibitem{Altmannshofer:2008dz}
W.~Altmannshofer, P.~Ball, A.~Bharucha, A.~J. Buras, D.~M. Straub and M.~Wick,
  \emph{{Symmetries and Asymmetries of $B \to K^{*} \mu^{+} \mu^{-}$ Decays in
  the Standard Model and Beyond}},
  \href{https://doi.org/10.1088/1126-6708/2009/01/019}{\emph{JHEP} {\bfseries
  01} (2009) 019}, [\href{https://arxiv.org/abs/0811.1214}{{\ttfamily
  0811.1214}}].

\bibitem{Buras:1990fn}
A.~J. Buras, M.~Jamin and P.~H. Weisz, \emph{{Leading and Next-to-leading {QCD}
  Corrections to $\epsilon$ Parameter and $B^0 - \bar{B}^0$ Mixing in the
  Presence of a Heavy Top Quark}},
  \href{https://doi.org/10.1016/0550-3213(90)90373-L}{\emph{Nucl. Phys. B}
  {\bfseries 347} (1990) 491--536}.

\bibitem{Urban:1997gw}
J.~Urban, F.~Krauss, U.~Jentschura and G.~Soff, \emph{{Next-to-leading order
  QCD corrections for the B0 anti-B0 mixing with an extended Higgs sector}},
  \href{https://doi.org/10.1016/S0550-3213(98)00043-1}{\emph{Nucl. Phys. B}
  {\bfseries 523} (1998) 40--58},
  [\href{https://arxiv.org/abs/hep-ph/9710245}{{\ttfamily hep-ph/9710245}}].

\bibitem{Lenz:2019lvd}
A.~Lenz and G.~Tetlalmatzi-Xolocotzi, \emph{{Model-independent bounds on new
  physics effects in non-leptonic tree-level decays of B-mesons}},
  \href{https://doi.org/10.1007/JHEP07(2020)177}{\emph{JHEP} {\bfseries 07}
  (2020) 177}, [\href{https://arxiv.org/abs/1912.07621}{{\ttfamily
  1912.07621}}].

\bibitem{Buras:2012fs}
A.~J. Buras and J.~Girrbach, \emph{{Complete NLO QCD Corrections for Tree Level
  Delta F = 2 FCNC Processes}},
  \href{https://doi.org/10.1007/JHEP03(2012)052}{\emph{JHEP} {\bfseries 03}
  (2012) 052}, [\href{https://arxiv.org/abs/1201.1302}{{\ttfamily 1201.1302}}].

\bibitem{Buras:2001ra}
A.~J. Buras, S.~Jager and J.~Urban, \emph{{Master formulae for Delta F=2 NLO
  QCD factors in the standard model and beyond}},
  \href{https://doi.org/10.1016/S0550-3213(01)00207-3}{\emph{Nucl. Phys. B}
  {\bfseries 605} (2001) 600--624},
  [\href{https://arxiv.org/abs/hep-ph/0102316}{{\ttfamily hep-ph/0102316}}].

\bibitem{LHCb:2020lmf}
{\scshape LHCb} collaboration, R.~Aaij et~al., \emph{{Measurement of
  $CP$-Averaged Observables in the $B^{0}\rightarrow K^{*0}\mu^{+}\mu^{-}$
  Decay}}, \href{https://doi.org/10.1103/PhysRevLett.125.011802}{\emph{Phys.
  Rev. Lett.} {\bfseries 125} (2020) 011802},
  [\href{https://arxiv.org/abs/2003.04831}{{\ttfamily 2003.04831}}].

\bibitem{Descotes-Genon:2013vna}
S.~Descotes-Genon, T.~Hurth, J.~Matias and J.~Virto, \emph{{Optimizing the
  basis of $B\to K^*ll$ observables in the full kinematic range}},
  \href{https://doi.org/10.1007/JHEP05(2013)137}{\emph{JHEP} {\bfseries 05}
  (2013) 137}, [\href{https://arxiv.org/abs/1303.5794}{{\ttfamily 1303.5794}}].

\bibitem{Altmannshofer:2014pba}
W.~Altmannshofer, S.~Gori, M.~Pospelov and I.~Yavin, \emph{{Neutrino Trident
  Production: A Powerful Probe of New Physics with Neutrino Beams}},
  \href{https://doi.org/10.1103/PhysRevLett.113.091801}{\emph{Phys. Rev. Lett.}
  {\bfseries 113} (2014) 091801},
  [\href{https://arxiv.org/abs/1406.2332}{{\ttfamily 1406.2332}}].

\bibitem{Altmannshofer:2019zhy}
W.~Altmannshofer, S.~Gori, J.~Mart\'\i{}n-Albo, A.~Sousa and M.~Wallbank,
  \emph{{Neutrino Tridents at DUNE}},
  \href{https://doi.org/10.1103/PhysRevD.100.115029}{\emph{Phys. Rev. D}
  {\bfseries 100} (2019) 115029},
  [\href{https://arxiv.org/abs/1902.06765}{{\ttfamily 1902.06765}}].

\bibitem{Li:2019sty}
T.~Li, Q.-F. Xiang, Q.-S. Yan, X.~Zhang and H.~Zhou, \emph{{Isospin-violating
  dark matter in a U(1)' model inspired by $E_6$}},
  \href{https://doi.org/10.1103/PhysRevD.101.035016}{\emph{Phys. Rev. D}
  {\bfseries 101} (2020) 035016},
  [\href{https://arxiv.org/abs/1908.00423}{{\ttfamily 1908.00423}}].

\bibitem{Frandsen:2011cg}
M.~T. Frandsen, F.~Kahlhoefer, S.~Sarkar and K.~Schmidt-Hoberg, \emph{{Direct
  detection of dark matter in models with a light Z'}},
  \href{https://doi.org/10.1007/JHEP09(2011)128}{\emph{JHEP} {\bfseries 09}
  (2011) 128}, [\href{https://arxiv.org/abs/1107.2118}{{\ttfamily 1107.2118}}].

\bibitem{Li:2022qrl}
T.~Li, Q.~Xiang, X.~Yin and H.~Zhou, \emph{{Generic U(1)X models inspired from
  SO(10)}}, \href{https://doi.org/10.1103/PhysRevD.106.075010}{\emph{Phys. Rev.
  D} {\bfseries 106} (2022) 075010},
  [\href{https://arxiv.org/abs/2201.03878}{{\ttfamily 2201.03878}}].

\bibitem{Straub:2018kue}
D.~M. Straub, \emph{{flavio: a Python package for flavour and precision
  phenomenology in the Standard Model and beyond}},
  \href{https://arxiv.org/abs/1810.08132}{{\ttfamily 1810.08132}}.

\bibitem{Buras:2014fpa}
A.~J. Buras, J.~Girrbach-Noe, C.~Niehoff and D.~M. Straub, \emph{{$ B\to
  {K}^{\left(\ast \right)}\nu \overline{\nu} $ decays in the Standard Model and
  beyond}}, \href{https://doi.org/10.1007/JHEP02(2015)184}{\emph{JHEP}
  {\bfseries 02} (2015) 184},
  [\href{https://arxiv.org/abs/1409.4557}{{\ttfamily 1409.4557}}].

\bibitem{Bellini:2011rx}
G.~Bellini et~al., \emph{{Precision measurement of the 7Be solar neutrino
  interaction rate in Borexino}},
  \href{https://doi.org/10.1103/PhysRevLett.107.141302}{\emph{Phys. Rev. Lett.}
  {\bfseries 107} (2011) 141302},
  [\href{https://arxiv.org/abs/1104.1816}{{\ttfamily 1104.1816}}].

\bibitem{Borexino:2013zhu}
{\scshape Borexino} collaboration, G.~Bellini et~al., \emph{{Final results of
  Borexino Phase-I on low energy solar neutrino spectroscopy}},
  \href{https://doi.org/10.1103/PhysRevD.89.112007}{\emph{Phys. Rev. D}
  {\bfseries 89} (2014) 112007},
  [\href{https://arxiv.org/abs/1308.0443}{{\ttfamily 1308.0443}}].

\bibitem{Borexino:2017rsf}
{\scshape Borexino} collaboration, M.~Agostini et~al., \emph{{First
  Simultaneous Precision Spectroscopy of $pp$, $^7$Be, and $pep$ Solar
  Neutrinos with Borexino Phase-II}},
  \href{https://doi.org/10.1103/PhysRevD.100.082004}{\emph{Phys. Rev. D}
  {\bfseries 100} (2019) 082004},
  [\href{https://arxiv.org/abs/1707.09279}{{\ttfamily 1707.09279}}].

\bibitem{Lindner:2018kjo}
M.~Lindner, F.~S. Queiroz, W.~Rodejohann and X.-J. Xu, \emph{{Neutrino-electron
  scattering: general constraints on $Z^\prime$ and dark photon models}},
  \href{https://doi.org/10.1007/JHEP05(2018)098}{\emph{JHEP} {\bfseries 05}
  (2018) 098}, [\href{https://arxiv.org/abs/1803.00060}{{\ttfamily
  1803.00060}}].

\bibitem{Falkowski:2018dsl}
A.~Falkowski, S.~F. King, E.~Perdomo and M.~Pierre, \emph{{Flavourful $Z'$
  portal for vector-like neutrino Dark Matter and $R_{K^{(*)}}$}},
  \href{https://doi.org/10.1007/JHEP08(2018)061}{\emph{JHEP} {\bfseries 08}
  (2018) 061}, [\href{https://arxiv.org/abs/1803.04430}{{\ttfamily
  1803.04430}}].

\bibitem{Altmannshofer:2014cfa}
W.~Altmannshofer, S.~Gori, M.~Pospelov and I.~Yavin, \emph{{Quark flavor
  transitions in $L_\mu-L_\tau$ models}},
  \href{https://doi.org/10.1103/PhysRevD.89.095033}{\emph{Phys. Rev. D}
  {\bfseries 89} (2014) 095033},
  [\href{https://arxiv.org/abs/1403.1269}{{\ttfamily 1403.1269}}].

\bibitem{Morel:2020dww}
L.~Morel, Z.~Yao, P.~Clad\'e and S.~Guellati-Kh\'elifa, \emph{{Determination of
  the fine-structure constant with an accuracy of 81 parts per trillion}},
  \href{https://doi.org/10.1038/s41586-020-2964-7}{\emph{Nature} {\bfseries
  588} (2020) 61--65}.

\bibitem{Misiak:1992bc}
M.~Misiak, \emph{{The $b \to se^+ e^-$ and $b \to s\gamma$ decays with
  next-to-leading logarithmic QCD corrections}},
  \href{https://doi.org/10.1016/0550-3213(93)90235-H}{\emph{Nucl. Phys. B}
  {\bfseries 393} (1993) 23--45}.

\bibitem{Buras:1994dj}
A.~J. Buras and M.~Munz, \emph{{Effective Hamiltonian for B ---\ensuremath{>}
  X(s) e+ e- beyond leading logarithms in the NDR and HV schemes}},
  \href{https://doi.org/10.1103/PhysRevD.52.186}{\emph{Phys. Rev. D} {\bfseries
  52} (1995) 186--195}, [\href{https://arxiv.org/abs/hep-ph/9501281}{{\ttfamily
  hep-ph/9501281}}].

\bibitem{Buras:2003mk}
A.~J. Buras, A.~Poschenrieder, M.~Spranger and A.~Weiler, \emph{{The Impact of
  universal extra dimensions on B ---\ensuremath{>} X(s) gamma, B
  ---\ensuremath{>} X(s) gluon, B ---\ensuremath{>} X(s) mu+ mu-, K(L)
  ---\ensuremath{>} pi0 e+ e- and epsilon-prime / epsilon}},
  \href{https://doi.org/10.1016/j.nuclphysb.2003.11.010}{\emph{Nucl. Phys. B}
  {\bfseries 678} (2004) 455--490},
  [\href{https://arxiv.org/abs/hep-ph/0306158}{{\ttfamily hep-ph/0306158}}].

\bibitem{Buras:1994qa}
A.~J. Buras, M.~E. Lautenbacher, M.~Misiak and M.~Munz, \emph{{Direct CP
  violation in K(L) ---\ensuremath{>} pi0 e+ e- beyond leading logarithms}},
  \href{https://doi.org/10.1016/0550-3213(94)90138-4}{\emph{Nucl. Phys. B}
  {\bfseries 423} (1994) 349--383},
  [\href{https://arxiv.org/abs/hep-ph/9402347}{{\ttfamily hep-ph/9402347}}].

\bibitem{Bartsch:2009qp}
M.~Bartsch, M.~Beylich, G.~Buchalla and D.~N. Gao, \emph{{Precision Flavour
  Physics with $B \to K \nu \bar\nu$ and $B \to K l^+ l^-$}},
  \href{https://doi.org/10.1088/1126-6708/2009/11/011}{\emph{JHEP} {\bfseries
  11} (2009) 011}, [\href{https://arxiv.org/abs/0909.1512}{{\ttfamily
  0909.1512}}].

\bibitem{Charles:1998dr}
J.~Charles, A.~Le~Yaouanc, L.~Oliver, O.~Pene and J.~C. Raynal, \emph{{Heavy to
  light form-factors in the heavy mass to large energy limit of QCD}},
  \href{https://doi.org/10.1103/PhysRevD.60.014001}{\emph{Phys. Rev. D}
  {\bfseries 60} (1999) 014001},
  [\href{https://arxiv.org/abs/hep-ph/9812358}{{\ttfamily hep-ph/9812358}}].

\bibitem{Beneke:2000wa}
M.~Beneke and T.~Feldmann, \emph{{Symmetry breaking corrections to heavy to
  light B meson form-factors at large recoil}},
  \href{https://doi.org/10.1016/S0550-3213(00)00585-X}{\emph{Nucl. Phys. B}
  {\bfseries 592} (2001) 3--34},
  [\href{https://arxiv.org/abs/hep-ph/0008255}{{\ttfamily hep-ph/0008255}}].

\bibitem{Wise:1992hn}
M.~B. Wise, \emph{{Chiral perturbation theory for hadrons containing a heavy
  quark}}, \href{https://doi.org/10.1103/PhysRevD.45.R2188}{\emph{Phys. Rev. D}
  {\bfseries 45} (1992) R2188}.

\bibitem{Burdman:1992gh}
G.~Burdman and J.~F. Donoghue, \emph{{Union of chiral and heavy quark
  symmetries}}, \href{https://doi.org/10.1016/0370-2693(92)90068-F}{\emph{Phys.
  Lett. B} {\bfseries 280} (1992) 287--291}.

\bibitem{Falk:1993fr}
A.~F. Falk and B.~Grinstein, \emph{{Anti-B ---\ensuremath{>} Anti-K e+ e- in
  Chiral Perturbation Theory}},
  \href{https://doi.org/10.1016/0550-3213(94)90554-1}{\emph{Nucl. Phys. B}
  {\bfseries 416} (1994) 771--785},
  [\href{https://arxiv.org/abs/hep-ph/9306310}{{\ttfamily hep-ph/9306310}}].

\bibitem{Casalbuoni:1996pg}
R.~Casalbuoni, A.~Deandrea, N.~Di~Bartolomeo, R.~Gatto, F.~Feruglio and
  G.~Nardulli, \emph{{Phenomenology of heavy meson chiral Lagrangians}},
  \href{https://doi.org/10.1016/S0370-1573(96)00027-0}{\emph{Phys. Rept.}
  {\bfseries 281} (1997) 145--238},
  [\href{https://arxiv.org/abs/hep-ph/9605342}{{\ttfamily hep-ph/9605342}}].

\bibitem{Buchalla:1998mt}
G.~Buchalla and G.~Isidori, \emph{{Nonperturbative effects in $\bar B \to X_s
  l^+ l^-$ for large dilepton invariant mass}},
  \href{https://doi.org/10.1016/S0550-3213(98)00261-2}{\emph{Nucl. Phys. B}
  {\bfseries 525} (1998) 333--349},
  [\href{https://arxiv.org/abs/hep-ph/9801456}{{\ttfamily hep-ph/9801456}}].

\bibitem{Bharucha:2015bzk}
A.~Bharucha, D.~M. Straub and R.~Zwicky, \emph{{$B\to V\ell^+\ell^-$ in the
  Standard Model from light-cone sum rules}},
  \href{https://doi.org/10.1007/JHEP08(2016)098}{\emph{JHEP} {\bfseries 08}
  (2016) 098}, [\href{https://arxiv.org/abs/1503.05534}{{\ttfamily
  1503.05534}}].

\end{thebibliography}

\end{document}